\documentclass[fleqn,10pt]{wlscirep}
\usepackage[resetlabels]{multibib}
\newcites{supp}{Supplementary Information References}

\title{Abrupt \textit{p-n} junction using ionic gating at zero-bias in bilayer graphene}
\author[1]{Sameer Grover}
\author[1,2]{Anupama Joshi}
\author[2]{Ashwin Tulapurkar}
\author[1,*]{Mandar M. Deshmukh}
\affil[1]{Department of Condensed Matter Physics and Materials Science, Tata Institute of Fundamental Research, Homi Bhabha Road, Mumbai 400005, India}
\affil[2]{Department of Electrical Engineering, Indian Institute of Technology Bombay, Mumbai 400076, India}
\affil[*]{deshmukh@tifr.res.in}
\begin{document}

\newcommand{\figdir}{}
\newcommand{\suppfigdir}{}
\newcommand{\eps}{jpg}

\begin{abstract}
  Graphene is a promising candidate for optoelectronic
  applications. In this report, a double gated bilayer graphene FET
  has been made using a combination of electrostatic and electrolytic
  gating in order to form an abrupt \textit{p-n} junction. The
  presence of two Dirac peaks in the gating curve of the fabricated
  device confirms the formation of a \textit{p-n} junction. At low
  temperatures, when the electrolyte is frozen intentionally, the
  photovoltage exhibits a six-fold pattern indicative of the hot
  electron induced photothermoelectric effect that has also been seen
  in graphene \textit{p-n} junctions made using metallic gates. We
  have observed that the photovoltage increases with decreasing
  temperature indicating a dominant role of supercollision
  scattering. Our technique can also be extended to other 2D materials
  and to finer features that will lead to \textit{p-n} junctions which
  span a large area, like a superlattice, that can generate a larger
  photoresponse. Our work creating abrupt \textit{p-n} junctions is
  distinct from previous works that use a source--drain bias voltage
  with a single ionic gate creating a spatially graded \textit{p-n}
  junction.
\end{abstract}

\flushbottom
\makeatletter
\let\oldmaketitle\maketitle
\let\old@maketitle\@maketitle
\let\old@thanks\@thanks
\let\old@author\@author
\let\old@date\@date
\let\old@title\@title
\let\oldtitle\title
\let\oldauthor\author
\let\olddate\date
\let\oldand\and
\makeatother
\maketitle
\thispagestyle{empty}

\section*{Introduction}


Graphene~\cite{Novoselov2004} has unique optical~\cite{Koppens2014,
  Bonaccorso2010, Bao2012} and electronic~\cite{CastroNeto2009}
properties which has made it a promising material for optoelectronic
devices such as photodetectors. The creation of \textit{p-n} junctions
with tunable chemical potentials allows the transduction of light into
electrical signals. In conventional electrostatic gating, a metallic
gate separated by a dielectric is used. The maximum
carrier density, typically $5\times10^{12}$ $\textrm{cm}^{-2}$ for
silicon dioxide, is limited by the breakdown strength of the
dielectric.


Electrolytic gating~\cite{Das2008} is an alternate technique that has
the advantage of large achievable carrier densities,
$\sim 4 \times 10^{13}$ $\textrm{cm}^{-2}$. This is limited by the
leakage current through the electrolyte ~\cite{Petach2014}. The large
capacitance is a result of the formation of an interfacial electrical
double layer~\cite{Shimotani2006} with a thickness of $\sim 1$
nm. Electrolytic gating has been used for tuning the carrier density
in various semiconductors such as organic
polymers~\cite{Vanmaekelbergh2007}, carbon
nanotubes~\cite{Rosenblatt2002}, and superconductors
~\cite{Ueno2008,Bollinger2011}.  Other non-tunable techniques for
large doping density include chemical
doping~\cite{Farmer2009,Peters2010}, exposure to electron
beams~\cite{Yu2015} or light\cite{Seo2014, Kim2013}, incorporating
dopant atoms in the lattice~\cite{Lin2016,Yan2012} and other growth
techniques~\cite{Wang2015}. Graphene can be electrolytically gated
using different types of electrolytes such as aqueous solutions of
salts \cite{Huang2015}, solid polymer electrolytes of lithium or
potassium salts in a polymer matrix~\cite{Das2008}, ionic
liquids~\cite{Chen2009} and their gels~\cite{Kim2010}.  Electrolytic
gating has been used previously in order to explore properties such as
superconductivity in MoS$_2$~\cite{Ye2012}. The achievement of large
carrier density is important for the study of properties of graphene
below the Bloch-Gr\"uneisen temperature~\cite{Efetov2010}. The added
advantage of this technique is that the ionic liquids have larger
optical transmission compared to a metallic top gate.

\begin{figure}[tb]
\centering
\includegraphics[width=160mm]{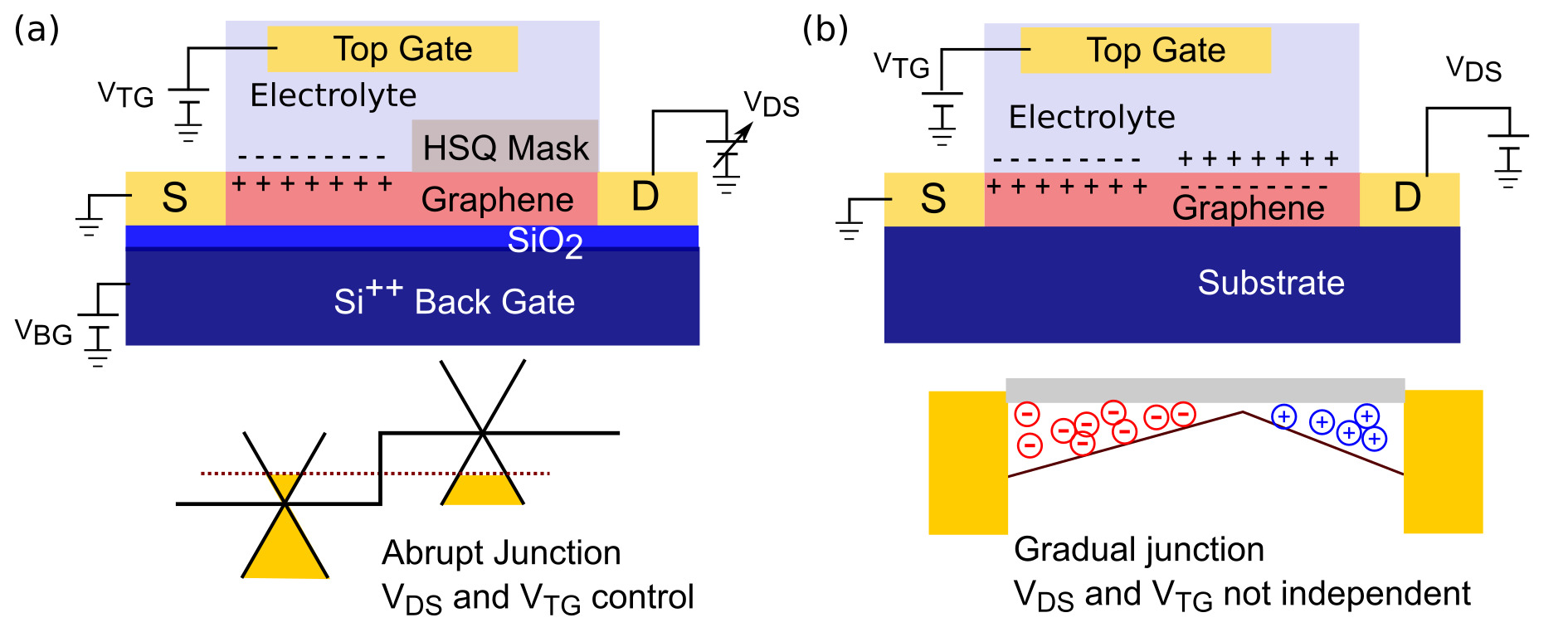}
\caption{\label{fig:intro} Comparison of the gating scheme discussed
  in this work with that used previously. \textbf{(a)}: Schematic of
  our gating scheme and the corresponding band diagram with abrupt
  \textit{p-n} junction and independent control of
  $V_{DS}$. \textbf{(b)}: Electrolytic gating scheme used in previous
  experiments~\cite{Chakraborty2009,Zhang2013} where a combination of
  $V_{DS}$ and $V_{GS}$ is used to create a junction which is not
  abrupt and does not allow independent control of $V_{DS}$ and the
  barrier height.}
\end{figure}

In this article, we demonstrate the formation of a \textit{p-n}
junction in graphene using a combination of electrostatic and
electrolytic gating at zero bias. Part of the graphene is covered with
a protective layer of hydrogen silsesquioxane (HSQ) resist which
prevents the electrolytic top gate from influencing the entire
graphene region.  Creating \textit{p-n} junctions at zero bias using
electrolyte gates is scalable and has not been done before. Graphene
\textit{p-n} junctions have previously been created using metallic
gates and with electrolytic gating by putting ionic liquid drop on
millimetre sized CVD graphene~\cite{He2014} and by applying a
drain-source bias comparable with the gate
voltage~\cite{Chakraborty2009}. Using the existing non-zero bias
technique, an abrupt \textit{p-n} junction is difficult to
realize. With our technique, we can create an abrupt profile which is
important for optoelectronic applications. Further, we have
independent control of the source drain bias voltage and \textit{p-n}
junction barrier height and using a zero bias is
possible. Table~\ref{table:comparison} summarizes work related to the
formation of \textit{p-n} junction by various techniques and
establishes novel aspect of our work. Figure~\ref{fig:intro} shows a
comparison of the abrupt \textit{p-n} junction profile that we have
formed and compared it to previous reports that use a source--drain
bias voltage to create a gradual junction using electrolytic gating.

\begin{table}[ht]
\centering
\begin{tabular}{|l|l|p{0.4\textwidth}|}
\hline
\textbf{Reference} & \textbf{Material Used} & \textbf{Study} \\
\hline
Zhang~\textit{et. al.}~\cite{Zhang2013} & MoS$_2$ & 
Spatially graded \textit{p-n} junction with bias dependent barrier height using ionic liquid\\
\hline
Chakraborty~\textit{et. al.}~\cite{Chakraborty2009} & Bilayer Graphene & 
Spatially graded \textit{p-n} junction with bias dependent barrier height using solid polymer electrolyte \\
\hline
He~\textit{et. al.}~\cite{He2014} & Graphene (CVD) & 
\textit{p-n-p} junction using ionic liquid and electrostatic back gate. 
Junction is abrupt but each region is a few millimetres wide and spatial control of the geometry is not possible.\\
\hline
Our Work & Bilayer graphene & Hybrid gating (combination of electrostatic and electrolytic gating) using ionic liquid, formation of abrupt junction and photoresponse study.\\
\hline
\end{tabular}
\caption{Comparison of reports  of formation of \textit{p-n} junction in graphene and related 2-D materials}
\label{table:comparison}
\end{table}

To demonstrate the \textit{p-n} junction we measure the gating curve
and observe two Dirac peaks. We also study the electrical properties
of the \textit{p-n} junction and the photoresponse as a function of
the junction barrier height and temperature. We find that the
photoresponse is dominated by the photothermoelectric effect,
characterized by a sixfold pattern in the photovoltage, similar to the
results obtained with electrostatic dual gates~\cite{Gabor2011}.  The
photovoltage increases as the temperature decreases which is
indicative of hot electron thermalization by disorder assisted
supercollisions~\cite{Song2012}.

\section*{Experiment}

\begin{figure}[tb]
\centering
\includegraphics[width=150mm]{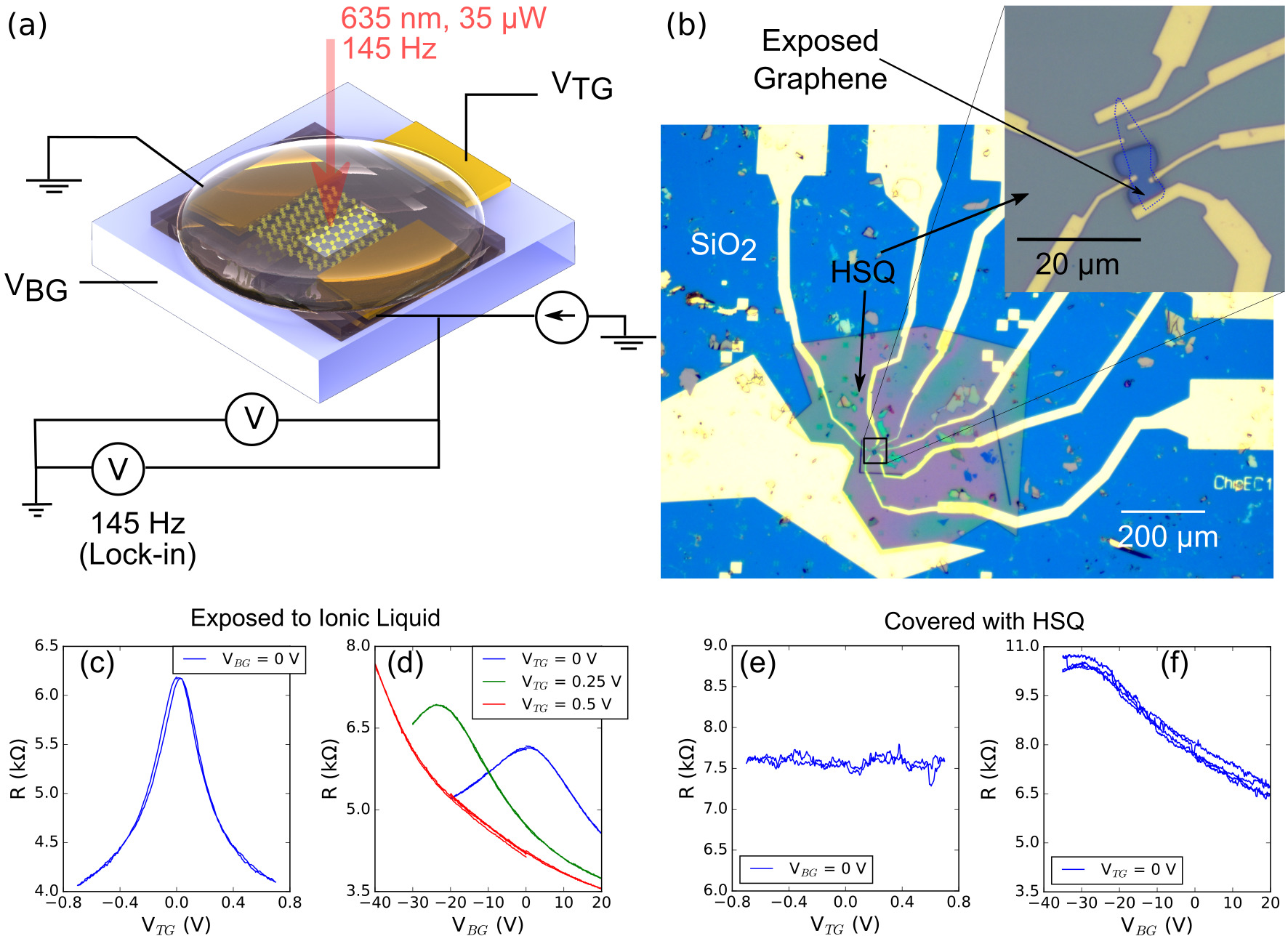}
\caption{\label{fig:electrical} Optical image and resistance
  measurements. \textbf{(a)}: Schematic of a graphene device which is
  gated both electrostatically and electochemcically and the
  measurement circuit with simultaneous measurement of resistance and
  photovoltage. (Supplementary Section VIII). \textbf{(b)}: Optical
  image of the graphene device on which measurements have been
  performed. The substrate is silicon with 300 nm of silicon
  dioxide. Source and drain electrodes, and a top gate electrode which
  is large in comparison with graphene, are made from titanium and
  gold. The graphene flake is partially masked with 500 nm thick
  HSQ. The ionic liquid drop is small and does not touch the aluminium
  wire bonds.  \textbf{(c)}, \textbf{(d)}, \textbf{(e)}, \textbf{(f)}:
  Resistance measurements as a function of top gate and back gate
  voltages at 280 K measured separately for the exposed and covered
  regions. \textbf{(c)} and \textbf{(d)}: Resistance variation with
  the top gate and back gate in the region exposed to the ionic
  liquid. This region is influenced by both the gates. \textbf{(e)}
  and \textbf{(f)}: Resistance variation with the top and back gates
  in the region covered with HSQ. The absence of gating with the top
  gate shows that HSQ is able to shield part of the graphene flake.}
\end{figure}

Graphene flakes identified with visual contrast and Raman spectroscopy
(Supplementary Section I) were mechanically exfoliated on Si/SiO$_2$
(300~nm) chips and metallization with titanium (7~nm) and gold (85~nm)
is done using standard electron beam lithography. The ionic liquid
reacts with electrodes made from chromium and gold and this limits the
metals that can be used. A large in-plane electrode is made in the
vicinity of graphene to serve as the top gate electrode.

In order to mask the graphene to protect it from the ionic liquid, we
have tried partial coverage with polymethylmethacrelate (PMMA) resist,
with overexposed PMMA, and HSQ resist. We found that PMMA tended to
dissolve in the ionic liquid over time and LiClO$_4$ in a PEO matrix
was not sufficiently optically transparent (Supplementary Section II).
Multiple electrodes are present on each of the regions so that the the
electrical properties can be separately measured and we can verify
that the resist protects graphene from ionic gating. We have used a
thick (500 nm) HSQ protective layer patterned with a dose of 350
$\mu$C/cm$^2$. We have also fabricated devices completely covered with
HSQ in order to verify that they are not affected by the ionic top
gate (Supplementary Section III).

An optical image of the device used and the measurement scheme is shown
in Figure~\ref{fig:electrical}. Measurements are performed in an
optical cryostat with simultaneous measurement of the resistance and
photovoltage.

\begin{figure}[tb]
\centering
\includegraphics[width=120mm]{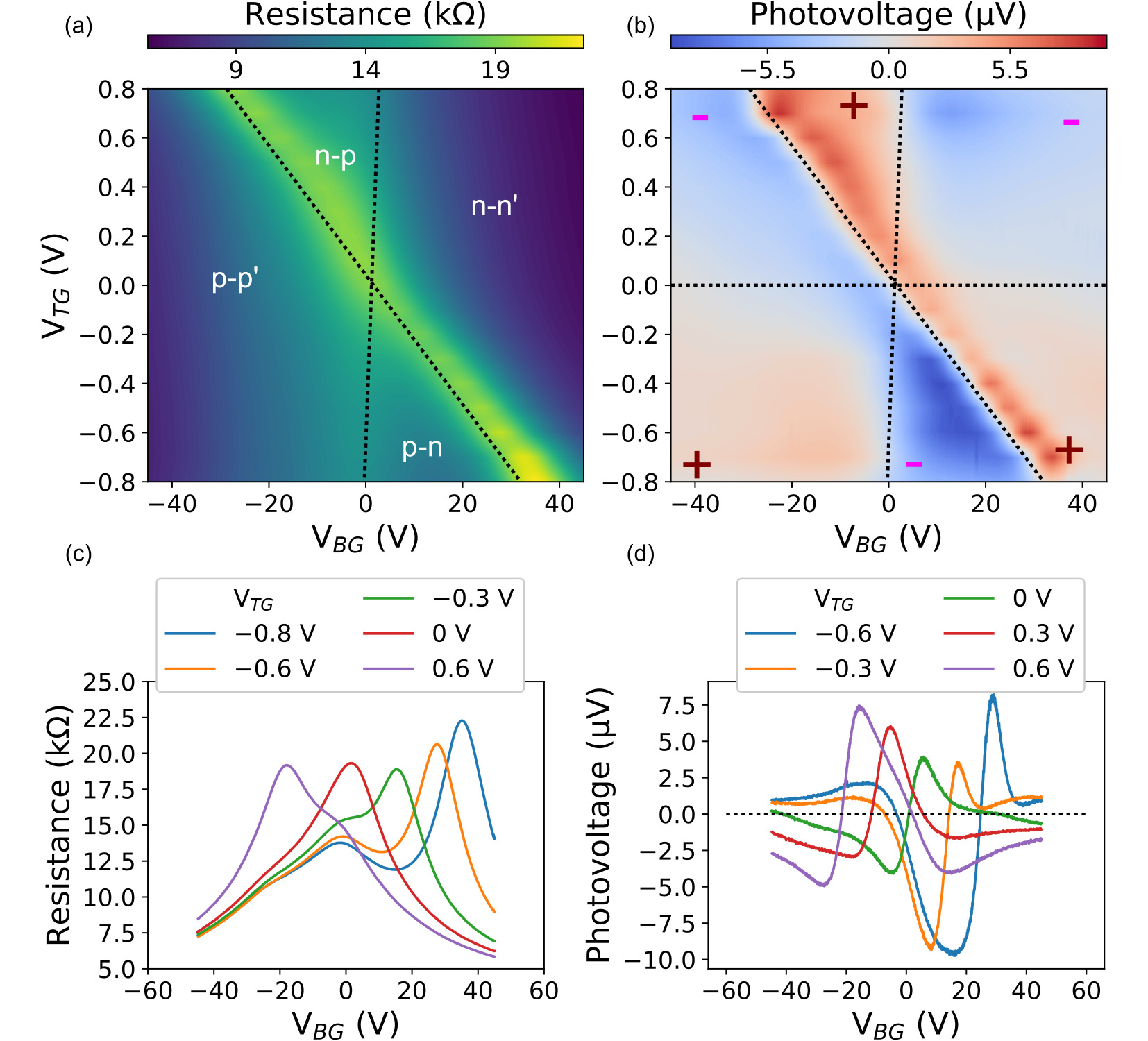}
\caption{\label{fig:lowtemp} Photovoltage and resistance measurements
  at 120~K. To change the top gate voltage, the device is warmed to
  273~K and the top gate voltage is adjusted and allowed to stabilize
  for half an hour before the device is cooled down to 120K.
  \textbf{(a)} The measured resistance as a function of both
  gates. The dotted lines indicate the charge neutrality peaks of the
  two regions and the polarity of the exposed and covered regions is
  indicated. \textbf{(b)} The corresponding photovoltage at the
  junction measured simultaneously with the resistance.  There is a
  clear signature of photothermoelectric effect indicated by the
  sixfold pattern~\cite{Gabor2011}. Six regions with alternating
  positive and negative signs are indicated by plus and minus signs
  and dotted lines that separate the six regions have been
  drawn. \textbf{(c)} Slices of the resistance data in
  Figure~\ref{fig:lowtemp}(a) exhibit two Dirac peaks. \textbf{(d)}
  Slices of the photovoltage with multiple zero crossings.}
\end{figure}

Ionic liquids~\cite{Galinski2006} are salts that are liquid at room
temperature, are transparent and stable. We have used the ionic liquid
EMI-Im (also called EMI-TFSI). Its glass transition
temperature~\cite{McEwen1999}is 175 K and melting point~\cite{Ngo2000}
is 258 K. Below the freezing point, the ions in ionic liquid are
immobilized and do not respond to externally applied fields. Our
measurements are conducted above (273 K) and below (30 - 150 K) the
freezing point. Changing the voltage applied to the ionic gate at
$T\le 150\: \text{K}$ is done by cycling the temperature.  Ionic
liquids are hygroscopic~\cite{Welton1999} and their electrical
properties are degraded by water absorption. Dehydration is usually
done by heating in vacuum or by freeze drying~\cite{Wellens2013}. Our
measurements are performed in an optical cryostat, and the device is
initially cooled to around 30 K in vacuum, leading to removal of
water.

We have noticed that it is necessary to put a very small amount of
ionic liquid so that the size of the droplet is small. When the ionic
liquid drop covers a large area, the back gate capacitance increases,
similar to the effect seen by Xia et al.~\cite{Xia2010}. We have
estimated that the effective increase in back gate capacitance is by
factor of $\sim$ 7 for a drop size of $\sim 1$ mm$^2$ (Supplementary
Section IV). Besides this, a large ionic liquid drop can touch the
aluminium wire bonds with which it reacts. We use a 25 $\mu m$
wire-bonder wire to wick a small amount of the ionic liquid and put in
on the graphene device. The HSQ surface, in contrast with overexposed
PMMA tends to repel the ionic liquid and this can be an added factor
in the ionic liquid preferentially covering only the exposed parts of
graphene. The refractive index of the ionic liquid and HSQ are nearly
identical ($\sim 1.5$) so that a surface of uniform refractive index
is presented to the incident light (Supplementary Figure S5).

We have also monitored the current through the top gate using a
sourcemeter to find the usable electrical limits of the top gate
voltage (Supplementary Section V). A top gate voltage of $\pm 2.5$ V results in a
current less than 1 nA. This current does not change if we turn on the
laser illumination, indicating an absence of photochemical
reactions. After applying a top gate voltage of 4 V, we have observed
that the Ti/Au electrodes directly under the ionic liquid were
corroded and the ones protected by HSQ were intact, further indicating
that HSQ is able to effectively shield the underlying material from
the ionic liquid (Supplementary Figure S5).

Figure~\ref{fig:electrical} shows resistance measurements at 280 K as
a function of the two gates for each of the exposed and masked region
separately. The region which is exposed to the ionic liquid shows
gating with both the top and back gates. We have calculated the ionic
gate capacitance using the back gate capacitance of 11.5 $nF/cm^2$ and
the capacitance ratio $\eta = C_{TG}/C_{BG} \sim 90$; this gives the
ionic gate capacitance $C_{TG} = 1 \mu F/cm^2$. We observe that the
resistance of the region of the flake which is masked by HSQ remains
unaffected by the top gate. All subsequent measurements are done
across the \textit{p-n} junction.

The variation of the resistance across the \textit{p-n} junction at
273 K for all values of $|V_{TG}|\le 0.8$~V and $V_{BG} \le 25$~V is
shown in Supplementary Section VI and indicates the formation of
\textit{p-n} junction at room temperature. Similar electrical
measurements performed at 120 K are shown in Figure~\ref{fig:lowtemp}.
These consist of four regions corresponding to \textit{p-n},
\textit{n-p}, \textit{n-n} and \textit{p-p} doping. The dotted lines
indicate the charge neutrality peaks of the two regions and the
polarity of the exposed and covered regions is indicated. The maximum
carrier density was determined by using the known value of the ionic
gate capacitance and the maximum gate voltage applied and it is found
to be $7 \times 10^{12}$ cm$^{-2}$.

To change the top gate voltage, the device was first warmed to 273~K
and the top gate voltage is adjusted and allowed to stabilize for 30
minutes before the device is cooled down to 120K. The photovoltage
generated at 120 K at the \textit{p-n} junction is given in
Figure~\ref{fig:lowtemp} along with line slices of the data for given
values of the top gate voltage. The data has been acquired by repeated
warming and cooling for each value of the top gate voltage. The
junction is robust and does not show any thermal cycling
effects (Supplementary Section IX).


The plot of the photovoltage at the junction shown in
Figure~\ref{fig:lowtemp}(b) as a function of the top and back gate
voltages exhibits a six-fold pattern. The photovoltage is zero along
the lines $n_{TG}=0$, $n_{BG}=0$ and $n_{TG}=n_{BG}$ and alternatively
positive and negative on either side ($n_{TG}$ and $n_{BG}$ are the
carrier densities in the region exposed to the ionic liquid and
covered with HSQ respectively). This manifests itself as two
intersections of the photoresponse curve with zero when measuring the
photovoltage as a function of the each gate, a shown in
Figure~\ref{fig:lowtemp}(c). The conventional mechanism of
photoresponse generation in semiconductors is the generation of an
electron hole pair due to the the absorption of a photon and their
separation under the influence of a built-in electric field such as
that arising at a \textit{p-n} junction. This is referred to as the
photovoltaic effect and the net photoresponse depends on the magnitude
and direction of band bending.

\begin{figure}
\centering
\includegraphics[width=153mm]{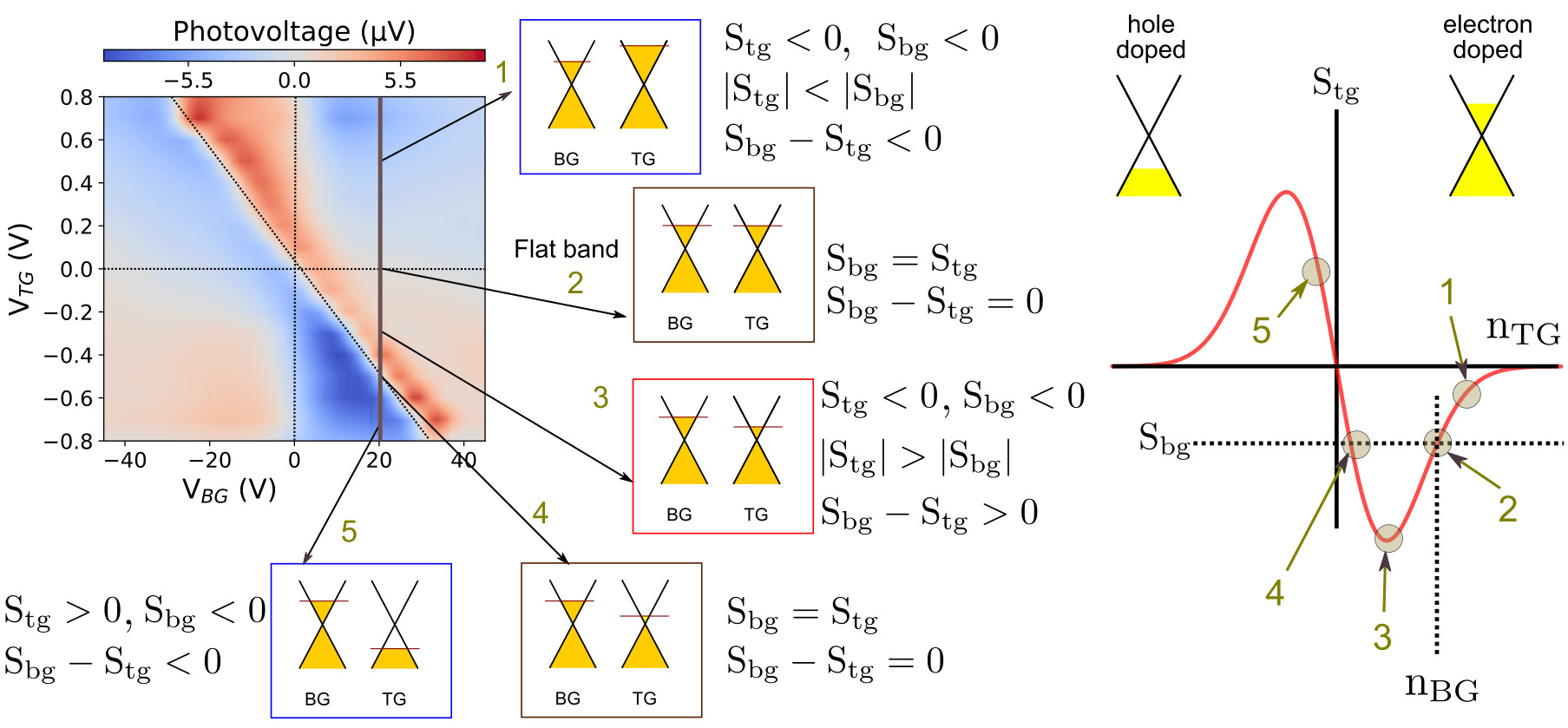}
\caption{\label{fig:seebeck2} The magnitude and signs of the Seebeck
  coefficient that leads to the six-fold pattern seen in the
  photovoltage data. Along the constant $V_{BG}$ line, the charge
  density in the covered region is constant and the exposed regions
  changes. The photovoltage is zero at point 2 and 4 where Seebeck
  coefficients are equal. The photovoltaic response is zero at point
  2, which is the flat band condition where the Fermi levels in both
  regions are equal. At point 4, the photoresponse is zero despite
  unequal Fermi levels and provides evidence of the
  photothermoelectric effect. The plot on the right schematically
  shows the variation of $S_{tg}$ with $n_{tg}$ at fixed values of
  $S_{bg}$ and $n_{bg}$. The existence of two distinct values of
  $n_{tg}$ where $S_{tg} = S_{bg}$ leads to the two photoresponse
  zeros seen in Figure~\ref{fig:lowtemp}(d).}
\end{figure}

We schematically illustrate in Figure~\ref{fig:seebeck2} that the
photovoltaic effect cannot explain the multiple polarity reversals
that are exhibited by the \textit{p-n} junction. Along the highlighted
line shown in Figure~\ref{fig:seebeck2} with a constant value of
$V_{BG}$, the charge density in the region covered with HSQ remains
constant. Along this line, only the Fermi level of the exposed region
changes. This change is monotonic and the photoresponse will only
change sign once by going through zero at the flat band condition at
point 2. However, the experimentally measured photoresponse data goes
through zero twice, at point 2 and point 4, and therefore cannot be
explained by the photovoltaic effect.

The photothermoelectric effect is consistent with our observations of
a six-fold pattern. Under the effect of light illumination, the
temperature at the \textit{p-n} junction increases and the temperature
at the source and drain contacts remains at the bath temperature. The
difference in carrier densities on either side of the junction leads
to different Seebeck coefficients in the two regions. This results in
a net photo-induced voltage developing across the device.

The multiple polarity reversals arise because of the functional form
of the Seebeck coefficient's dependence on the charge density,
illustrated in Figure~\ref{fig:seebeck2}. The Seebeck coefficient is
positive for hole doping and negative for electron doping. It is an
odd function and goes through a maximum before decreasing at large
carrier densities. For a given value of the Seebeck coefficient of the
covered region influenced by the back gate, there are two distinct
carrier densities where Seebeck coefficient of the exposed top-gated
region will be equal to that of back gated region, resulting in a zero
photovoltage. Along the highlighted line Figure~\ref{fig:seebeck2},
the sign and magnitude of the Seebeck coefficient and the photovoltage
at five distinct points is highlighted.

The photovoltage is given by difference in the Seebeck coefficients in
the two regions:
\begin{equation}
PV(V_{BG}, V_{TG}) = [S(n_{BG}) - S(n_{TG})] \Delta T
\label{eq:difference}
\end{equation}
and the Mott relation relates the Seebeck coefficient and the charge
density:
\begin{equation}
S = \frac{\pi^2 k_B^2 T}{3e} \frac{d(lnR)}{dn} \frac{dn}{dE}
\label{eq:mott}
\end{equation}
For bilayer graphene~\cite{Chakraborty2009},
$ n = -sign(E_F) \alpha [\gamma_1|E_F| + E_F^2]$ and
$dn/dE = -\sqrt{\alpha^2\gamma^2 + 4|n|\alpha}$, where $n<0$($n>0$)
for electron (hole) doping, $\alpha=(\pi(\hbar v_f)^2)^{-1}$, and the
inter-layer hopping energy $\gamma_1 \sim 390$ meV. The resistance
data is fitted to the form
$R = L/W (e\mu)^{-1}\frac{1}{\sqrt{n^2+n_0^2}}$, from which we get
$ d(lnR)/dn = \frac{-n}{n^2 + n_0^2}$.

\begin{figure}
\centering
\includegraphics[width=80mm]{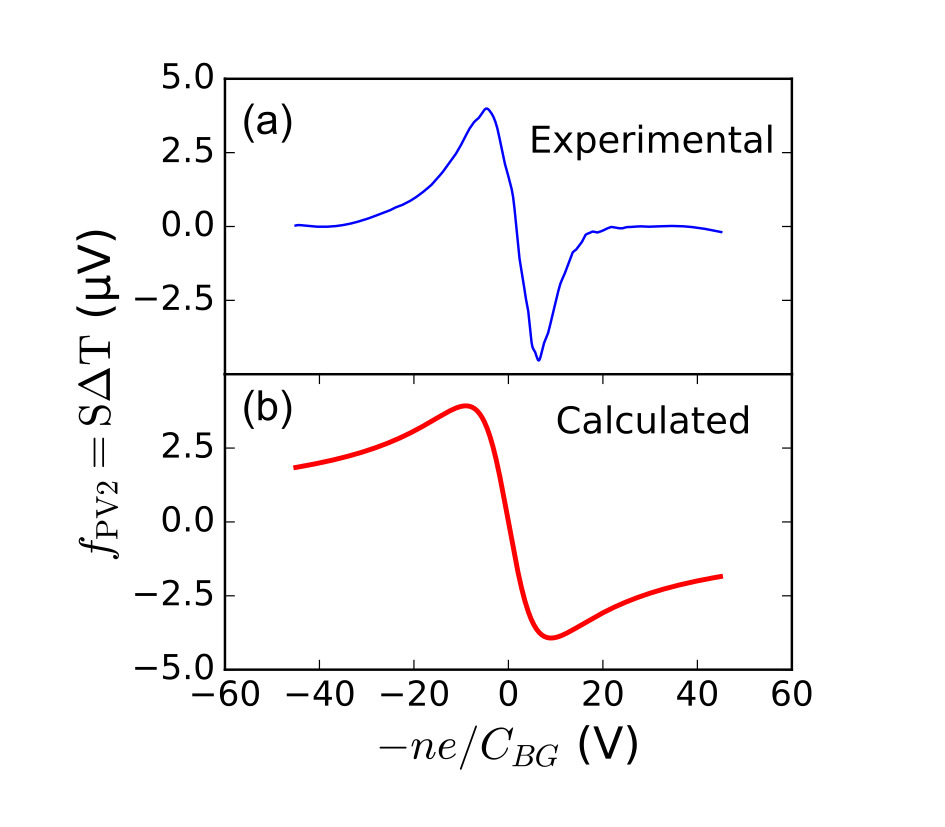}
\caption{\label{fig:seebeck1} A comparison of the
  product of the Seebeck coefficient and temperature increase obtained
  experimentally and calculated theoretically as a function of the
  carrier density. \textbf{(a)}: The experimental curve has been obtained by Fourier
  transforming the data in Figure~\ref{fig:lowtemp}(a) and represents
  the function $f_{PV2}$, which corresponds to the region exposed to
  the top gate. \textbf{(b)}: The theoretical curve has been calculated by
  calculating the Seebeck coefficient using the Mott relation and with
  a temperature increase of 50 mK.}
\end{figure}

We can estimate the value of the Seebeck coefficient from the
experimental data. The photovoltage arising form the
photothermoelectric can be written as a linear combination of two
functions that depend on the doping level in each region, similar to
equation~\ref{eq:difference}:
\begin{equation}
PV = f_{PV1}(V_{BG}) - f_{PV2}(V_{BG} + \eta (V_{TG}))
\end{equation}
where $\eta = C_{TG}/C_{BG}$ and we treat the functions $f_{PVi}$ as
unknowns and estimate them from the experimental data using two
dimensional Fourier transforms (Supplementary Section VII), similar to
the approach taken by Gabor \textit{et. al.}~\cite{Gabor2011}. The
function $f_{PV2}$ has been plotted in Figure~\ref{fig:seebeck1}(a) as
a function of the carrier density. For the photothermoelectric effect,
the function is related to the Seebeck coefficient as
$f_{PVi} = S\Delta T$ (S is the Seebeck coefficient and $\Delta T$ the
temperature increase). The experimental value is compared to the
theoretical result obtained using the Mott relation,
equation~\ref{eq:mott} and is shown in Figure~\ref{fig:seebeck1}(b).

The temperature increase is estimated from the experimental data as
$\Delta T = f_{PV}/S = 50$ mK. This is smaller than the expected
temperature increase of $\sim 1 K$ that would be expected from the
electronic thermal conductivity of graphene~\cite{Yigen2014} of
$\sim 2-5$ W/m/K, also obtainable from the the Wiedemann-Franz
relation. This difference is because the temperature increase of the
electronic subsystems obtained through the heat balance
equation~\cite{Song2012} is strongly dependent on the cooling
length. We have also neglected any light absorption in the ionic
liquid and the effective optical power reaching the device and the
corresponding temperature increase could be lower.

\begin{figure}
\centering
\includegraphics[width=140mm]{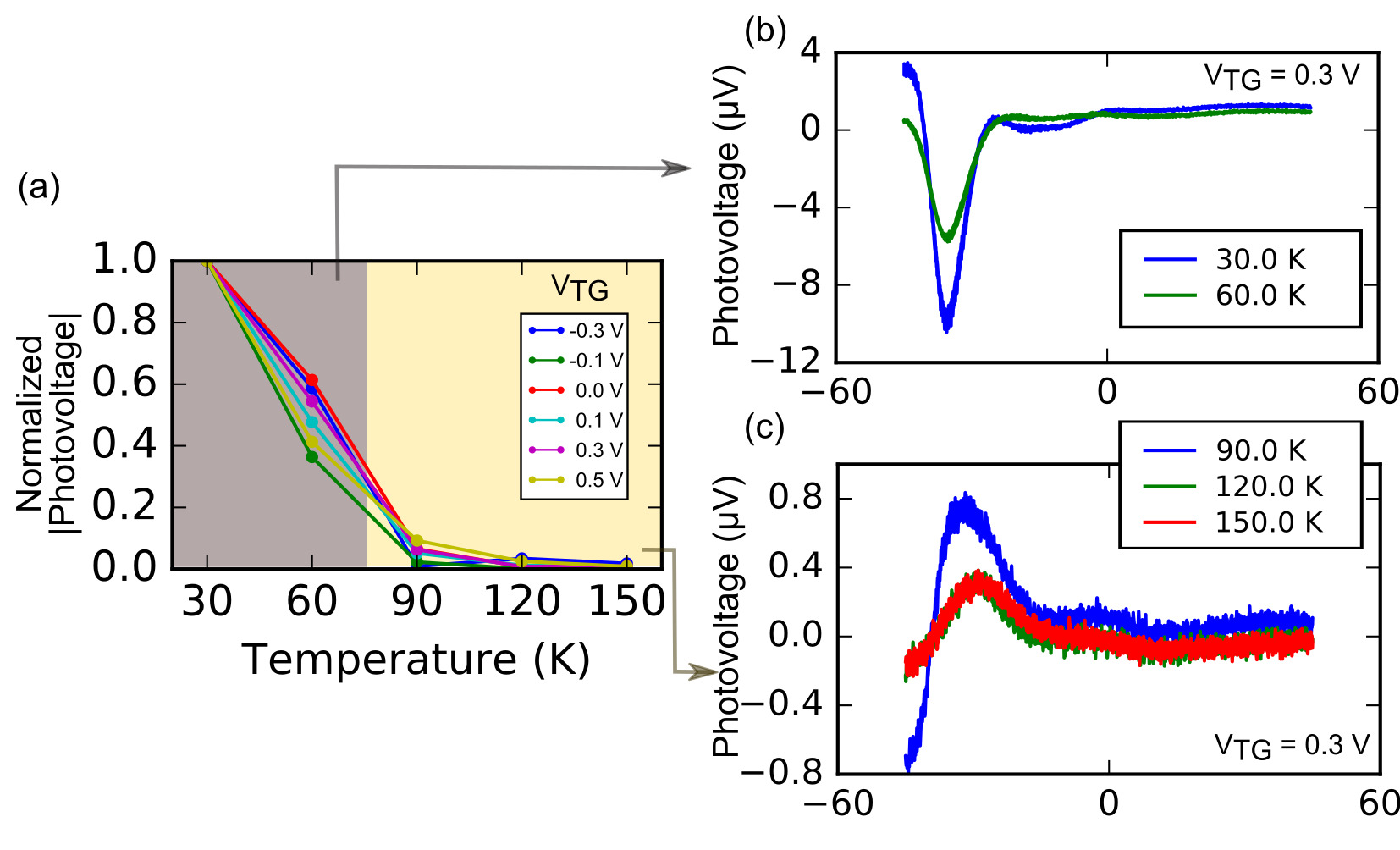}
\caption{\label{fig:tempdep} Temperature dependence of the
  photovoltage. \textbf{(a)} Normalized photovoltage magnitude as a
  function of temperature at different top gate voltages when
  $n_{BG} = -n_{TG}$.  The normalization factor depends on
  $V_{TG}$. \textbf{(b)} and \textbf{(c)}: The photovoltage as a
  function of the back gate at $V_{TG}=0.3$ V.  }
\end{figure}

Photovoltage measurements at temperatures ranging from 30 K to 150 K
have also been performed.  The normalized photovoltage magnitude at
different top gate voltages has been plotted in
Figure~\ref{fig:tempdep}(a). The photovoltage increases with
decreasing temperature. The photovoltage rises rapidly below 90 K and
in this range, the product of photovoltage and temperature is
constant, indicating that the photovoltage is inversely proportional
to temperature (Supplementary Section X). The photo-responsivities we
have obtained range from 100 mV/W at 273 K to 600 mV/W at 30 K.
Increase in the photovoltage at lower temperatures indicates an
increase in the cooling length, implying that the cooling rate
increases with temperature. This is similar to the trend
observed~\cite{Ma2014a} in monolayer graphene where this has been
attributed to disorder mediated supercollisions~\cite{Song2012,
  Graham2013,Betz2012} as the dominant electronic thermalization
mechanism.

In conclusion, we have achieved the formation of an abrupt
\textit{p-n} junction in graphene using a combination of electrostatic
and electrolytic gating using HSQ as a protective mask.  This
technique is scalable and the fabrication of an array of \textit{p-n}
junctions, such as a superlattice~\cite{Dubey2013}, can be realized by
the narrow features that can be lithographically patterned using
HSQ. The combination of larger optical transparency of ionic liquids
and the potential for larger carrier densities make this an
interesting system for optoelectronic studies.


\section*{Acknowledgements}

We acknowledge funding from the Department of Atomic Energy, the
Department of Science and Technology (Swarnajayanti Fellowship for
M.M.D) of the Government of India and ITC-PAC Grant
No. FA5209-15-P-0092.

\section*{Author contributions statement}
S.G. and A.J. performed the experiments. The manuscript was written
by S.G., A.J. and M.M.D. All authors discussed the results and commented on the manuscript.

\section*{Additional Information}
\textbf{Competing financial interests} The authors declare no competing financial interests.

\newpage
\makeatletter
\let\maketitle\oldmaketitle
\let\@maketitle\old@maketitle
\let\@thanks\old@thanks
\let\@author\old@author
\let\@date\old@date
\let\@title\old@title
\let\title\oldtitle
\let\author\oldauthor
\let\date\olddate
\let\and\oldand
\renewcommand{\@maketitle}{%
{%
\thispagestyle{empty}%
\vskip-36pt%
{\centering\sffamily\bfseries\fontsize{30}{35}\selectfont Supplementary Information\par}%
\vskip26pt%
{\raggedright\sffamily\bfseries\fontsize{20}{25}\selectfont \@title\par}%
\vskip10pt
{\raggedright\sffamily\fontsize{12}{16}\selectfont  \@author\par}
}%
}%
\makeatother

\title{Abrupt \textit{p-n} junction using ionic gating at zero-bias in bilayer graphene}

\maketitle

\makeatletter
\setcounter{figure}{0}
\renewcommand{\thefigure}{S\@arabic\c@figure}
\renewcommand{\theequation}{S\@arabic\c@equation}
\renewcommand{\thesection}{\@Roman\c@section}
\makeatother

\section{Raman spectrum of graphene}

\begin{figure}[ht]
  \centering
  \includegraphics[width=5in]{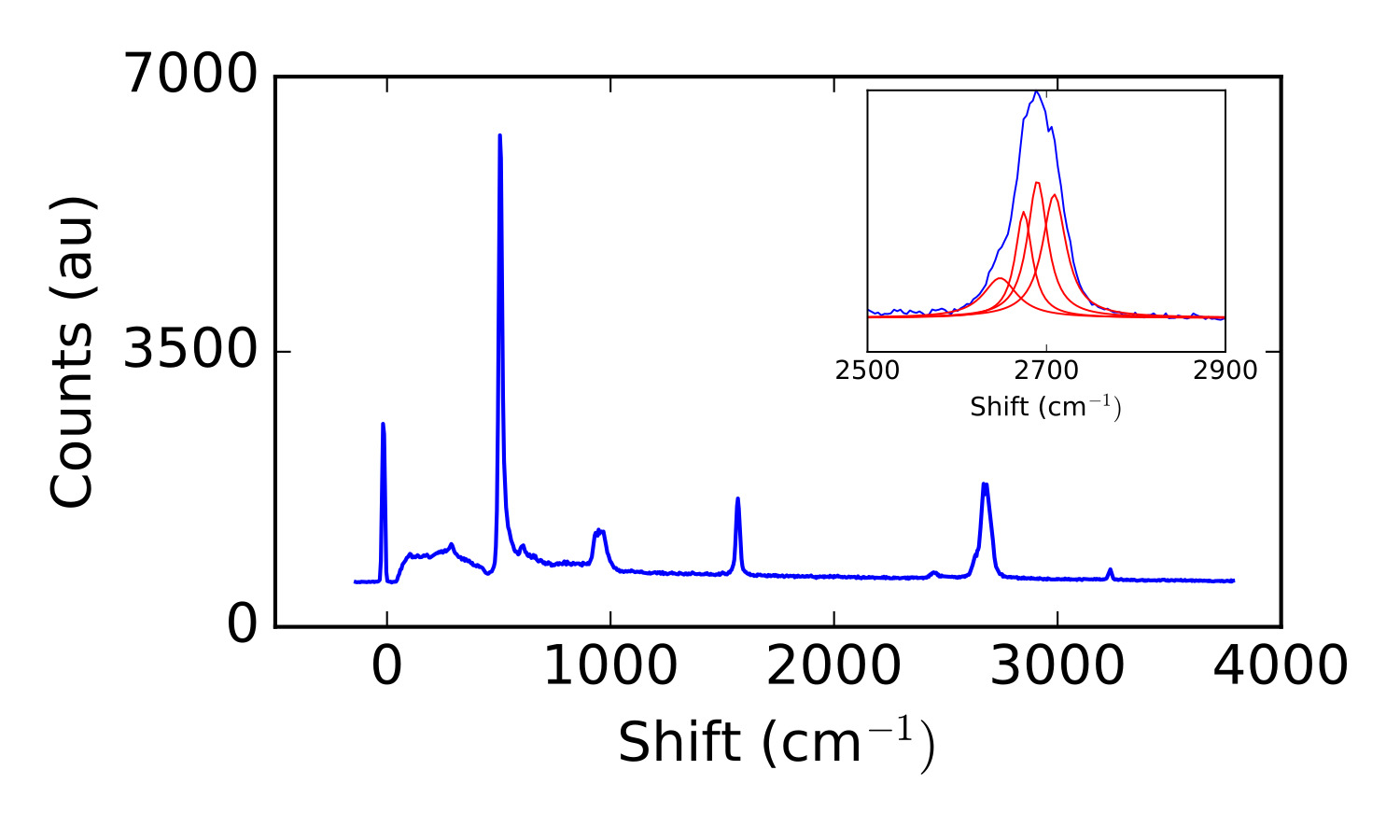}
  \caption{\label{fig:raman} Raman spectrum of bilayer graphene flake
    shown in Figure 1 in the main text}
\end{figure}

Figure~\ref{fig:raman} shows the Raman spectrum of the bilayer graphene
flake presented in the main text measured with an excitation
wavelength of 532 nm.

\clearpage

\section{Optical Transparency of LiClO$_4$/PEO}

We tried using a PEO/LiClO$_4$ solid electrolyte but found that it was
not optically transparent. Though it has been used before for optical
measurements~\citesupp{Das2008B}, it is difficult to apply it locally so
that it is uniform and does not contact the back gate.

\begin{figure}[h!]
    \centering
  \includegraphics[width=4in]{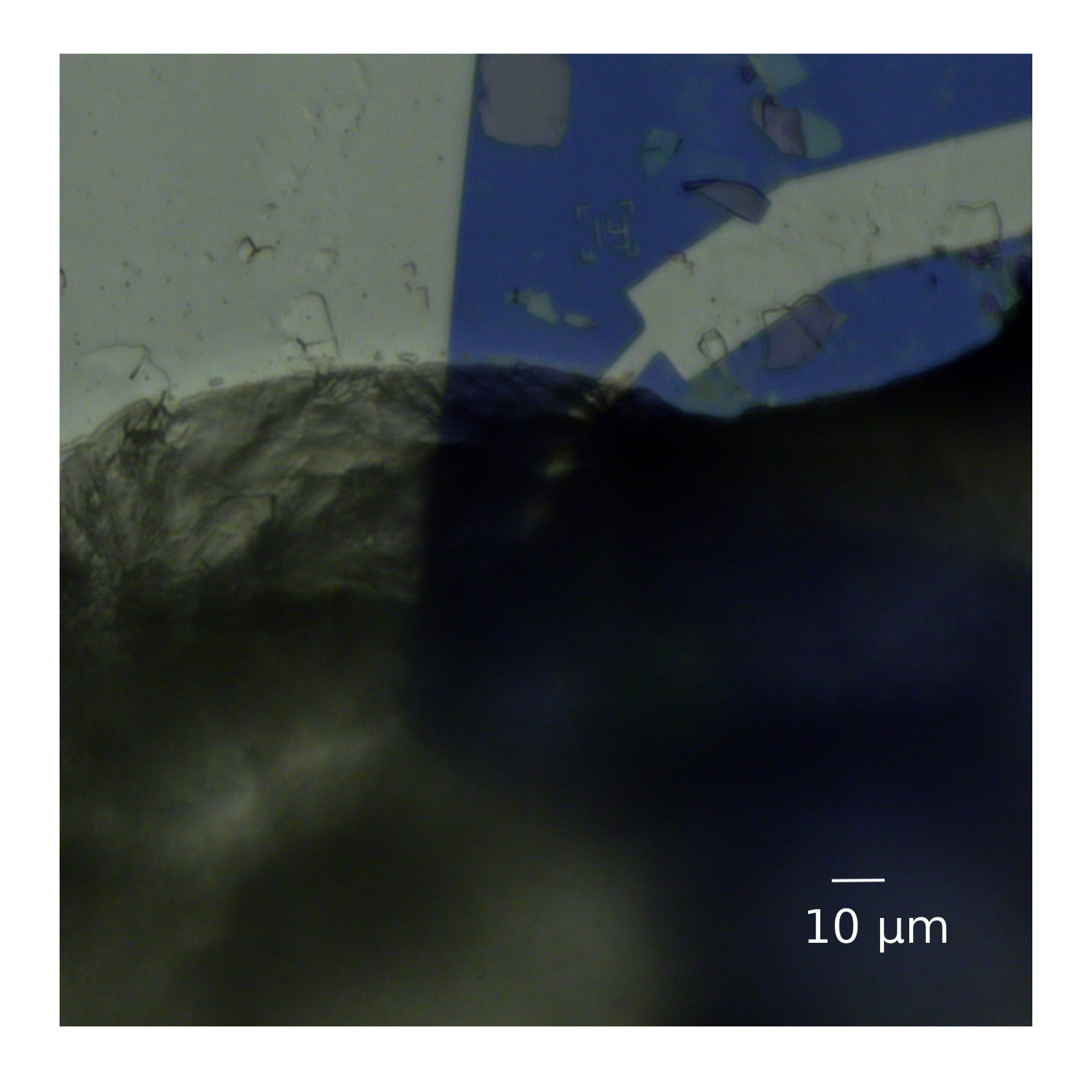}
  \caption{Optical image of a graphene device covered with a solid electrolyte made from LiClO$_4$/PEO.
    \label{fig:peo}}
\end{figure}

\clearpage

\section{Ionic Gate screening by HSQ and PMMA}

\begin{figure}[ht]
\centering
\includegraphics[width=6.5in]{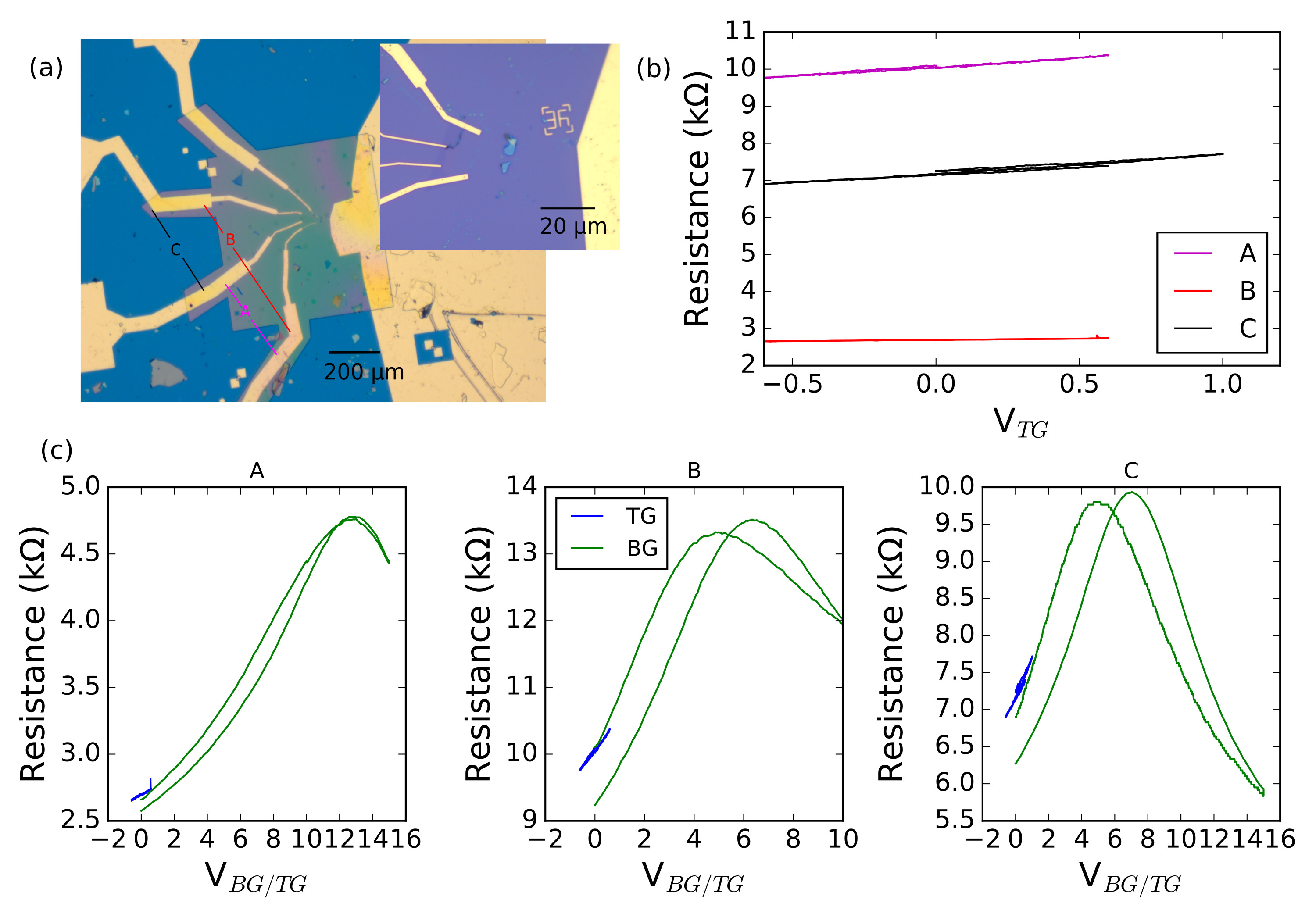}
\caption{(a) Optical image showing the graphene device completely
  covered with HSQ. (Inset) Optical image after metallization and
  before HSQ was deposited. Three pairs of electrodes on which
  measurements are done are labelled. (b) Variation of the device
  resistance as a function of the top gate voltage. (c) Gating with
  the top and back gate plotted on the same scale.
\label{fig:control}}
\end{figure}

In order to verify whether HSQ is able to effectively mask the ionic
liquid, we fabricated a graphene device which was completely covered
with HSQ. The measured resistance as a function of both the top gate
and back gate separately is given in Figure~\ref{fig:control}.

Even though the device is fully covered with HSQ,
Figure~\ref{fig:control}(b) shows that the change in the top gate
voltage causes the resistance to change slightly. This change is small
and indicative of a low capacitance. On plotting the gating curves of
the top gate and back gate on the same scale
(Figure~\ref{fig:control}(c)), it can be seen that the change in the
resistance because of the top gate is comparable to that due to the
back gate. This indicates that the capacitance between the
electrolytic top gate and graphene is smaller than what would be
expected from a few nanometres thick Debye layer and that the high
capacitance due to the electrolyte has been suppressed.

\begin{figure}[ht]
    \centering
\includegraphics[width=6.5in]{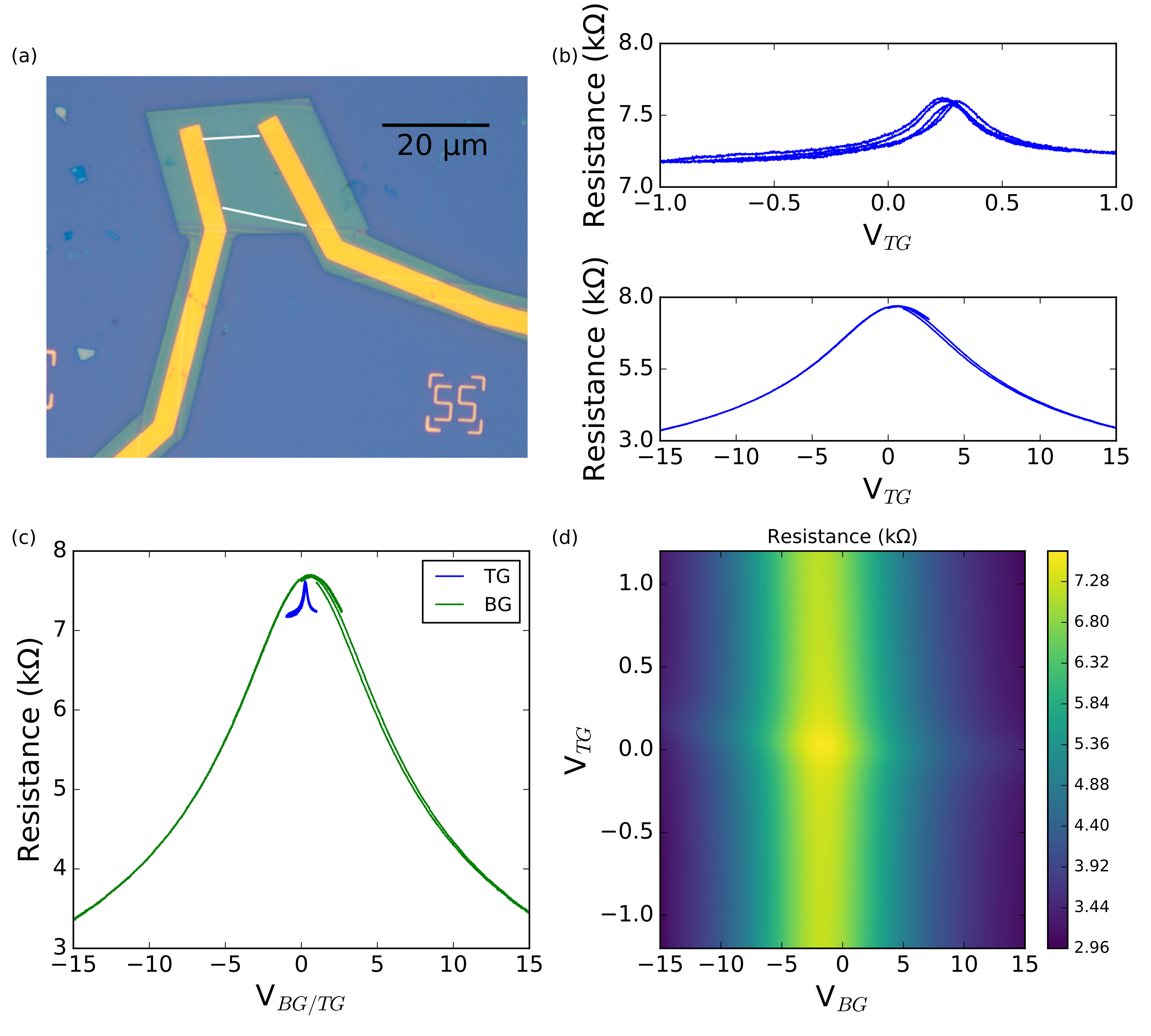}
\caption{(a) Optical image of the graphene device covered with PMMA. The boundaries of
  the flake are marked in blue. The entire region, including the
  electrodes are covered with overexposed PMMA. (b) The gating curves
  of the device with the top and back gates. (c) The two gating curves
  overlaid on each other. (d) The resistance as a function of top gate
  and back gate voltages.
  \label{fig:control_pmma}}
\end{figure}

We also made another device using PMMA overexposed at a dose of 10000
$\mu C/cm^2$ which completely covers the graphene flake. The gating
curves for this device are shown in Figure~\ref{fig:control_pmma}. The
resistance changes with the top gate and the slope $dG/dV_{TG}$ is
more than that of the back gate. However, the magnitude of the change
is small compared to that of the back gate. It is possible that there
are small areas within PMMA that the ionic liquid can percolate
through and affect graphene.

\begin{figure}[ht]
    \centering
  \includegraphics[width=6.5in]{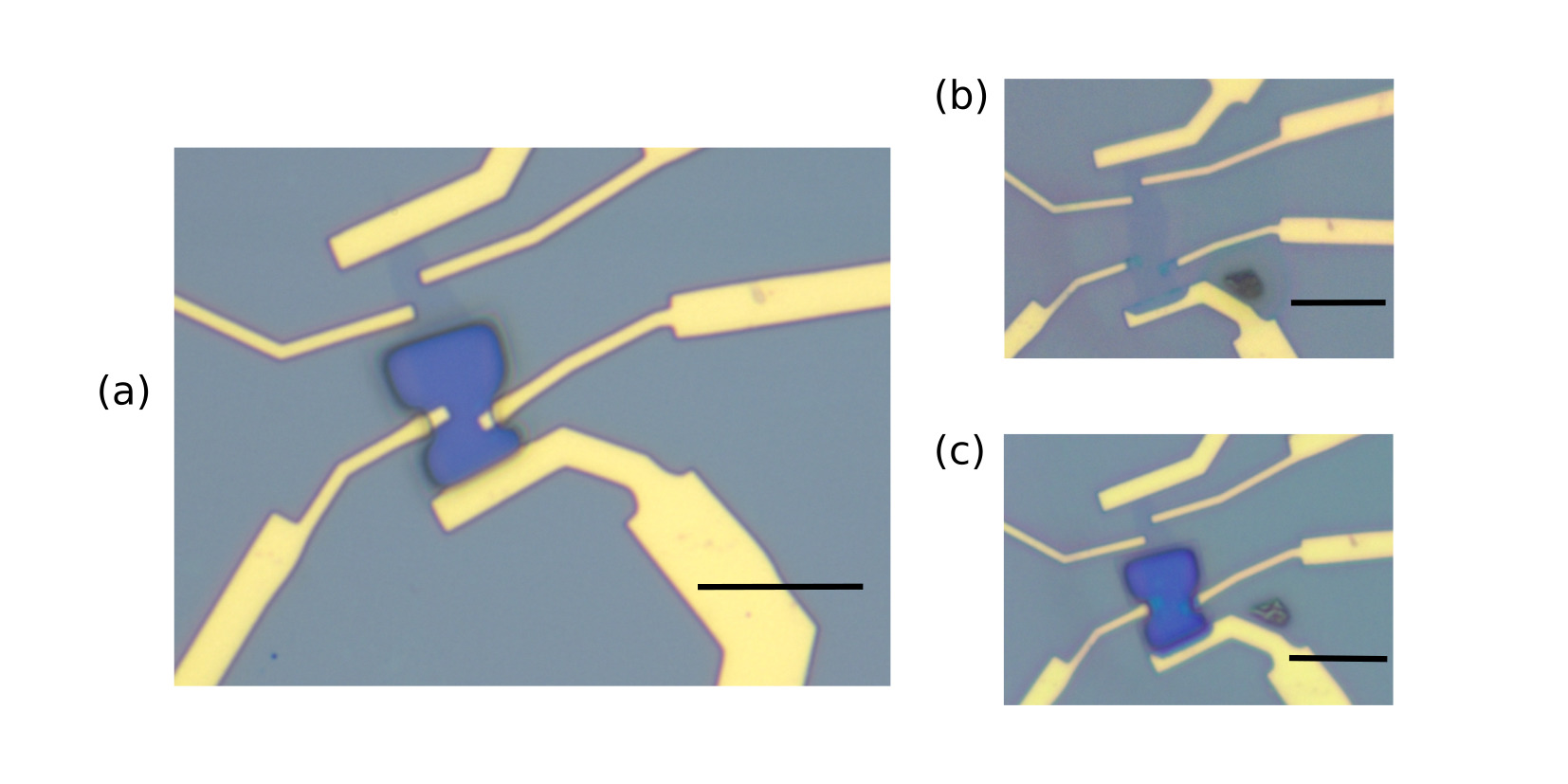}
  \caption{ Corrosion of electrodes caused by large top gate voltage
    (a) Optical image of device after HSQ patterning, before the ionic
    liquid was dropped. (b) Optical image after measurements with the
    ionic liquid still on the device. The boundary of HSQ is not
    visible because the refractive index of HSQ is identical to that of
    the ionic liquid. (c) Optical image after washing
    off the ionic liquid with isopropyl alcohol. Scale bar in all images is 10
    $\mu$m.
    \label{fig:corrosion}}
\end{figure}

We also applied large top gate voltages ($\sim 4.5$ V) to the bilayer
graphene device that is presented in the main text. This caused
corrosion of the electrodes. From the optical image given in
Figure~\ref{fig:corrosion}, we find that the electrodes covered by the
HSQ are unaffected, once again indicating that the HSQ is an effective
mask for the ionic liquid.

\clearpage

\section{Capacitance variation with area of ionic liquid}

\begin{figure}[ht]
  \centering
  \includegraphics[width=8cm]{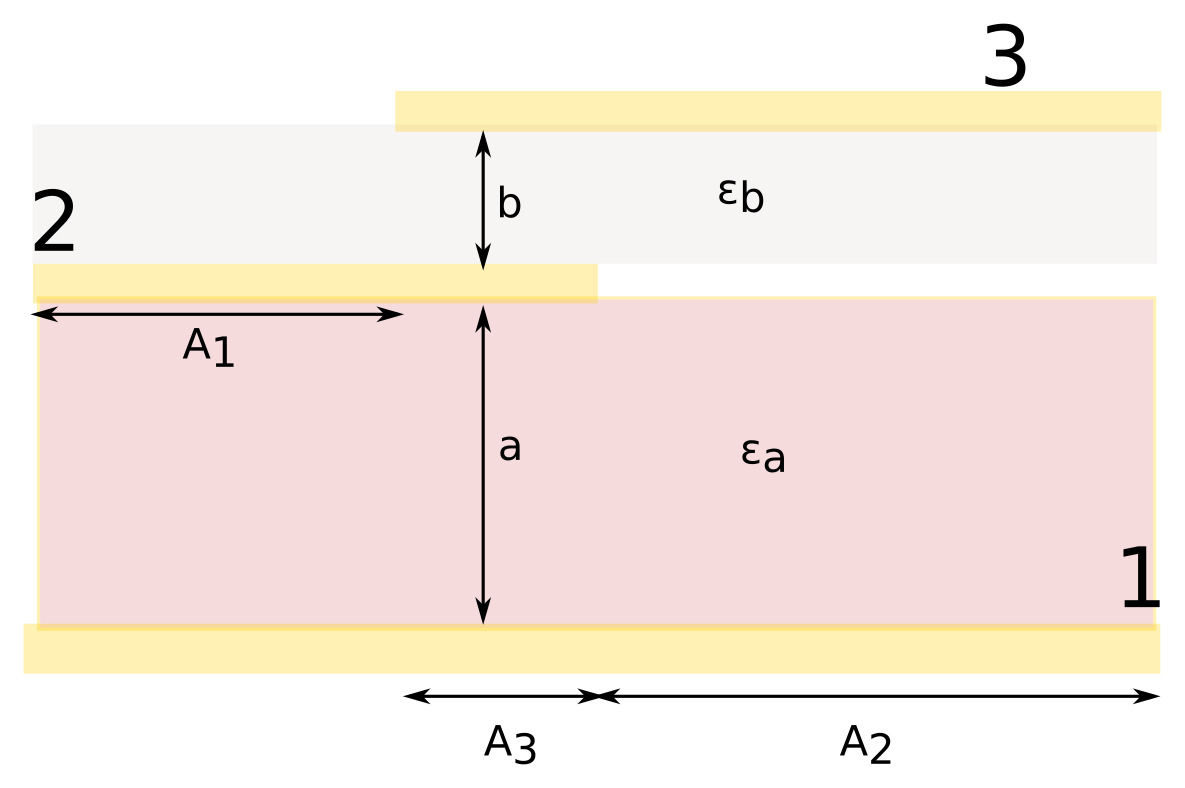}
\caption{\label{fig:cap1} Model for the dual gated FET that has been
  used in the calculations. The back gate, graphene and top gate are
  represented by the conductors 1,2 and 3 respectively. The bottom gate
  dielectric is SiO$_2$ and the electrical double layer is the top
  gate dielectric. The cross-sectional areas of overlap as indicated
  are $A_1$, $A_2$ and $A_3$}
\end{figure}

Ignoring the quantum capacitance by treating the graphene as a metal,
we model the device as a system of conductors as shown in
Figure~\ref{fig:cap1}. The back gate, graphene and top gate are denoted
by $1$, $2$ and $3$ and they overlap vertically with overlap areas
$A_i$ as indicated.

Neglecting fringing fields, we can write the capacitance matrix
relating the charge on each conductor $Q_i$ to its potential $V_i$ as:

\begin{equation}
  q_{1} =
  \Big[
    \frac{A_{2} \epsilon_{a} \epsilon_{b}}{a \epsilon_{b} + b
      \epsilon_{a}} + \frac{\epsilon_{a}}{a} \left(A_{1} +
    A_{3}\right)
    \Big] v_{1}
  +
  \Big[
  - \frac{\epsilon_{a}}{a} \left(A_{1} + A_{3}\right)
  \Big] v_{2}
    +
    \Big[
      - \frac{A_{2} \epsilon_{a} \epsilon_{b}}{a \epsilon_{b} + b
        \epsilon_{a}}
      \Big] v_{3}
\end{equation}

\begin{equation}
  q_{2} =
  \Big[
    - \frac{\epsilon_{a}}{a} \left(A_{1} + A_{3}\right)
    \Big] v_{1}
  +
  \Big[
 \left(\frac{A_{3} \epsilon_{b}}{b} + \frac{\epsilon_{a}}{a}
\left(A_{1} + A_{3}\right)\right)
\Big] v_{2}
  +
  \Big[
    - \frac{A_{3} \epsilon_{b}}{b} 
    \Big]v_{3}
\end{equation}

\begin{equation}
  q_{3} =
  \Big[  
  - \frac{A_{2} \epsilon_{a} \epsilon_{b}}{a \epsilon_{b} + b
    \epsilon_{a}}
    \Big] v_{1}
  +
    \Big[
      - \frac{A_{3} \epsilon_{b}}{b}
      \Big] v_{2}
    +
      \Big[
        \frac{A_{2} \epsilon_{a} \epsilon_{b}}{a \epsilon_{b} + b
          \epsilon_{a}} + \frac{A_{3} \epsilon_{b}}{b}
        \Big] v_{3}
\end{equation}

The effective capacitance per unit area between the back gate and graphene with the
top gate at a fixed potential is $C_{12} = q_1/v/f$ with $v_1 = v,
v_2=v_3=0$ with $f$ as the area on conductor 2 on which charge accumulates ($f \approx A_1 + A_3$).

\begin{equation}
  C_{12} = \frac{\epsilon_{a}}{a}\Bigg[\frac{1}{f \left(a \epsilon_{b} + b
    \epsilon_{a}\right)} \left(A_{1} a \epsilon_{b} + A_{1} b
  \epsilon_{a} + A_{2} a \epsilon_{b} + A_{3} a \epsilon_{b} + A_{3} b
  \epsilon_{a}\right)\Bigg]
\label{eq:final1}
\end{equation}

From the device geometry, we estimate the capacitance with $\epsilon_a
= 3.9 \epsilon_0$, $\epsilon_b = 16.5 \epsilon_0 $ ~\citesupp{Huang2011B},
$A_1 = (425\: \mu \textrm{m})^2$, $A_3 = (80 \: \mu \textrm{m})^2$, $b =
1$ $\textrm{nm}$, $ a = 300$ $ \textrm{nm}$, $f = (425 \:\mu
\textrm{m})^2$.

\begin{figure}[ht]
    \centering
  \includegraphics[width=14cm]{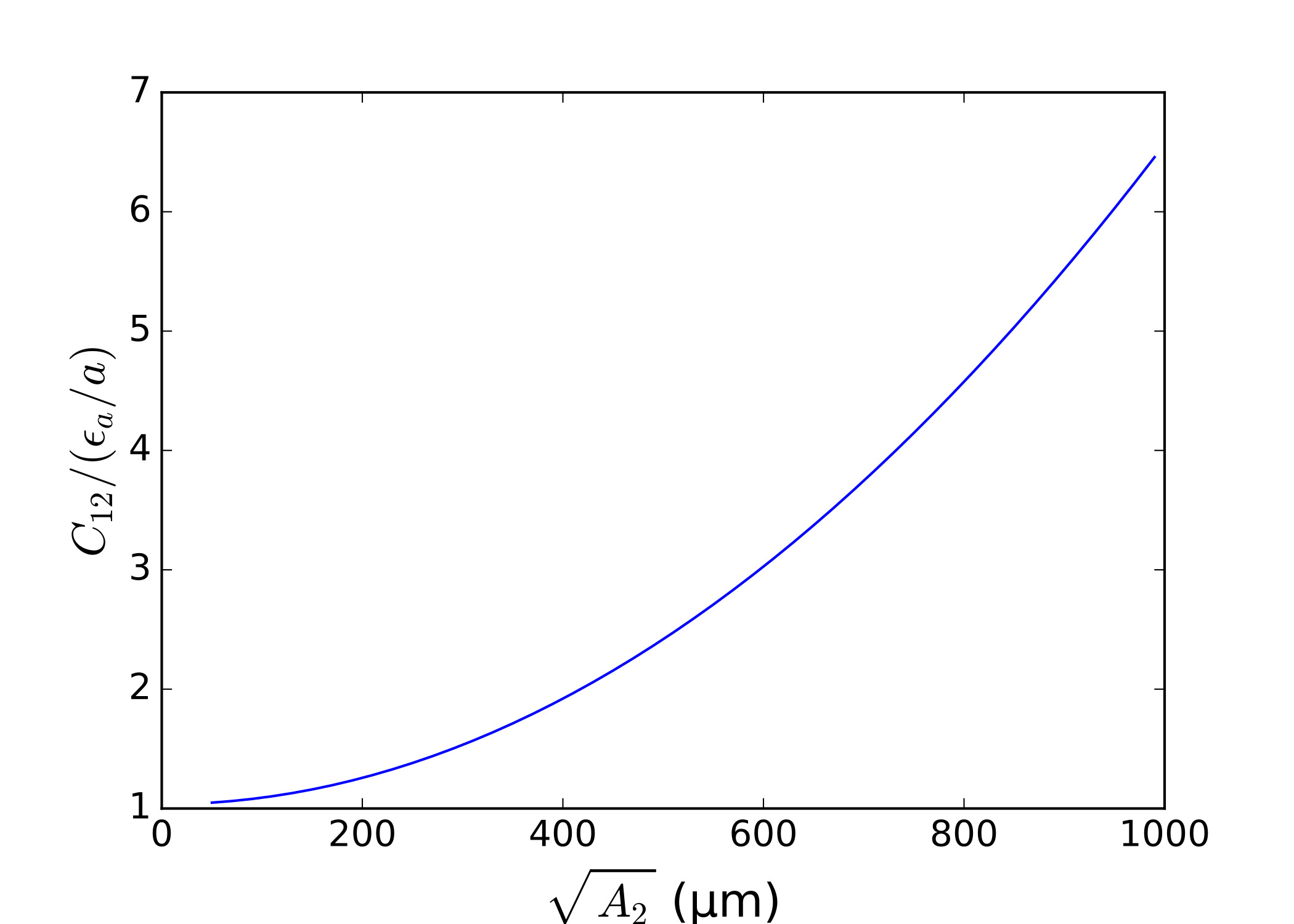}
  \caption{Numerical solution of equation~\ref{eq:final1}. The factor
    of the back gate capacitance increase is shown as a function of
    the ionic liquid drop lateral size.}
  \label{fig:cap2}
\end{figure}

Figure~\ref{fig:cap2} gives the factor of increase in the back gate
capacitance as a function of the ionic liquid drop dimension.

\clearpage

\section{Ionic top gate leakage current}

We have measured the leakage current of the ionic liquid by monitoring the
current drawn by the top gate. The current is less than 1 nA till
$V_{TG}=4$ V. However, we saw an abrupt change in the device
resistance at around 3.5 V

\begin{figure}[ht]
  \centering
  \includegraphics[width=5in]{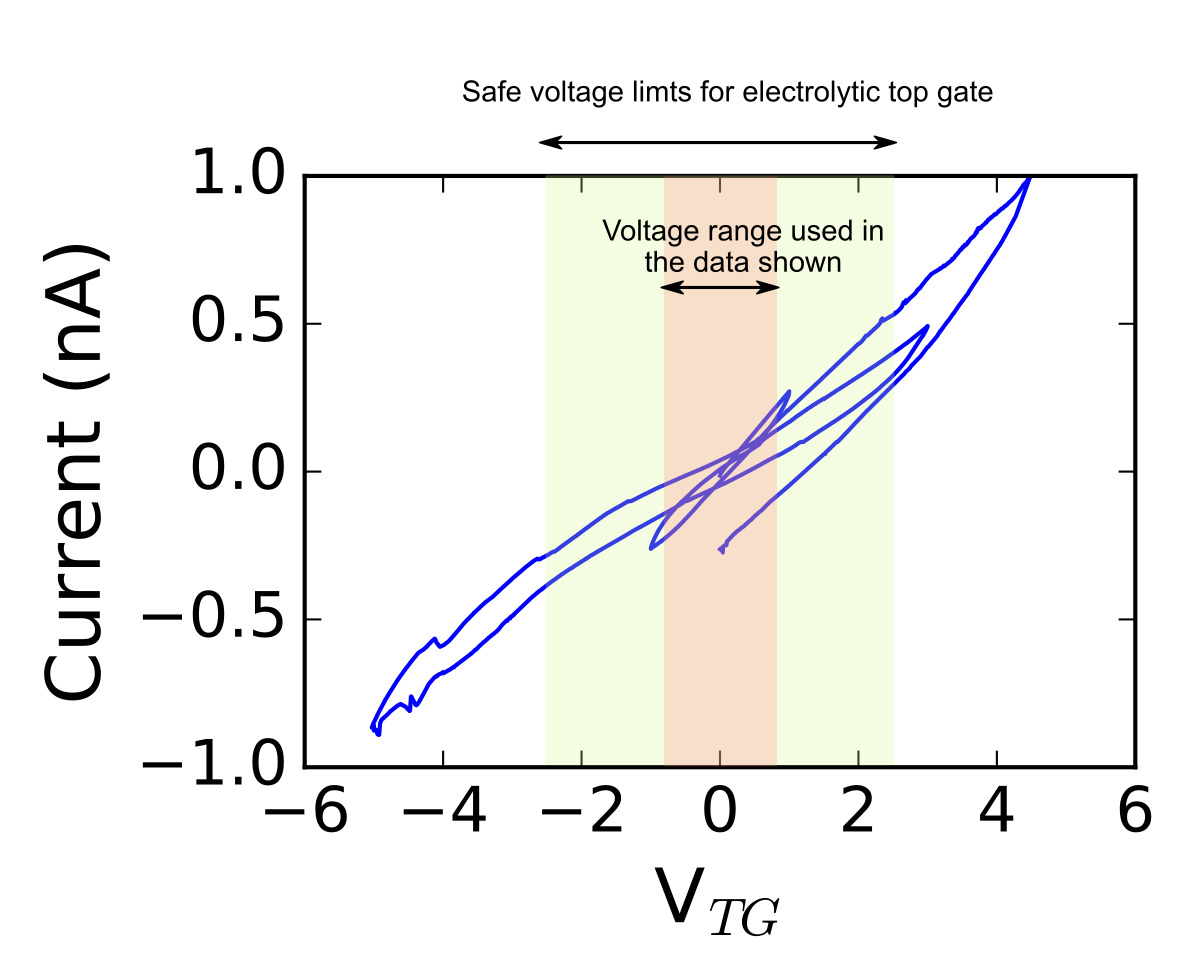}
  \caption{ The leakage current drawn by the ionic top gate as a function of the top
    gate voltage.
    \label{fig:tgcurrent}}
\end{figure}

\clearpage

\section{Electrical characterization of \textit{p-n} junction at room temperature}

\begin{figure}[ht]
  \centering
  \includegraphics[width=6in]{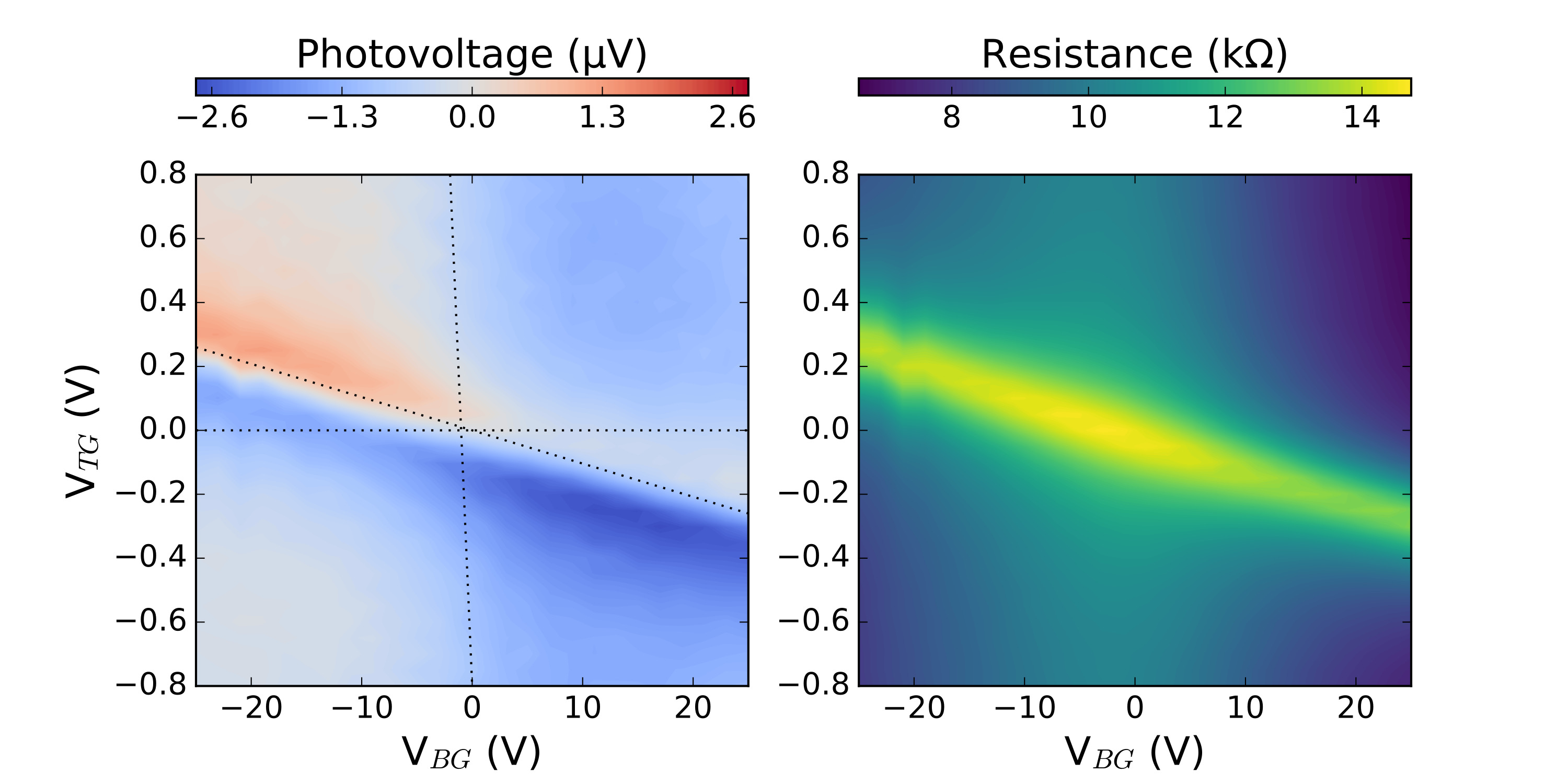}
  \caption{Photovoltage (left) and Resistance (right) of the device at 273 K.
    \label{fig:roomtemp}}
\end{figure}

Figure~\ref{fig:roomtemp} shows the resistance and corresponding
photovoltage at 273 K as a function of both the top and back gates.

\clearpage

\section{Fourier Transform Analysis}

If $z(x,y) = f(ax+b)$, $Fz(k_x, k_y) = Ff(k_x/a)\delta(ak_y -
bk_x)$, where $Fz$ and $Ff$ are the Fourier transforms of $z$ and $f$
respectively and $k=1/x$,

Accordingly, if the photovoltage $PV$ is given by:

\begin{equation}
PV(V_{BG},V_{TG}) = f_{PV1}(V_{BG}) - f_{PV2}(V_{BG} + \eta (V_{TG}))
\end{equation}

The Fourier transform of $f_{PV1}$ lies along the $1/V_{BG}$ axis and
$f_{PV2}$ lies along the $1/V_{TG} = \eta (1/V_{BG})$ axis. The values
of $f_{PVi}$, $i=1,2$ are found by masking with a Gaussian and taking
the inverse transform.

Since, for the data at 120K, acquisition of the photovoltage at each
top gate voltage involves heating and cooling, the data is sampled
coarsely at intervals of 0.1 V, with 17 points ranging from -0.8 V to
0.8 V, we have only shown the function form of $f_{PV2}$ in the main text.

\clearpage

\section{Measurement Circuit}

Light of wavelength 635 nm modulated at 145 Hz is focused on the
junction using a microscope objective. A current of 50 nA at 550 Hz is
applied across the source and drain to measure the resistance. Two
lock-in amplifiers, at 145 Hz and 550 Hz are used to measure the
photovoltage and the resistance respectively. (see Figure 1(a) in the main
text).

\clearpage

\section{Thermal cycling stability}

\begin{figure}
  \centering
\includegraphics[width=150mm]{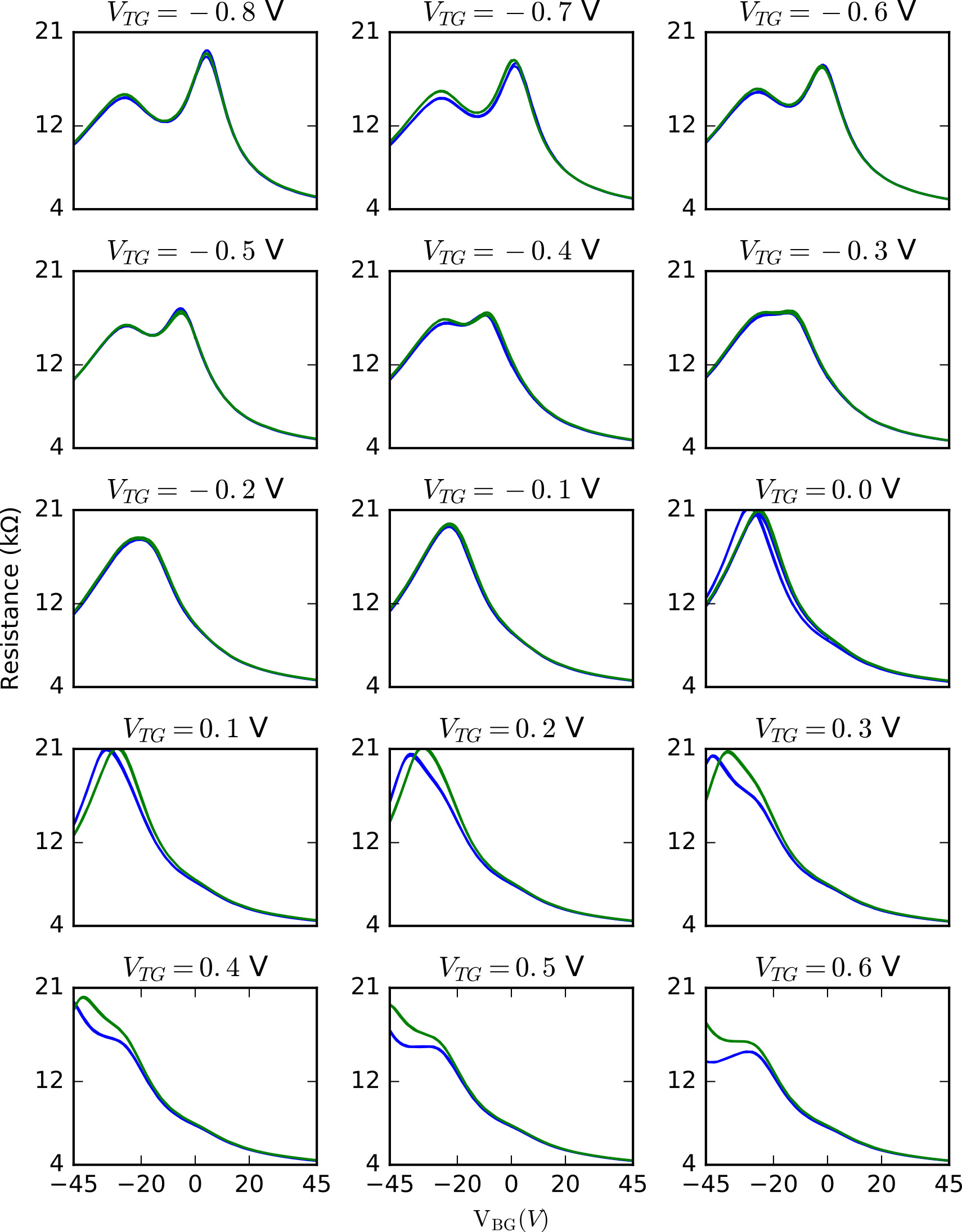}
\caption{Resistance at 120 K measured before and after 15 thermal
  cycling events.}
\label{fig:thermal1}
\end{figure}

\begin{figure}
    \centering
\includegraphics[width=150mm]{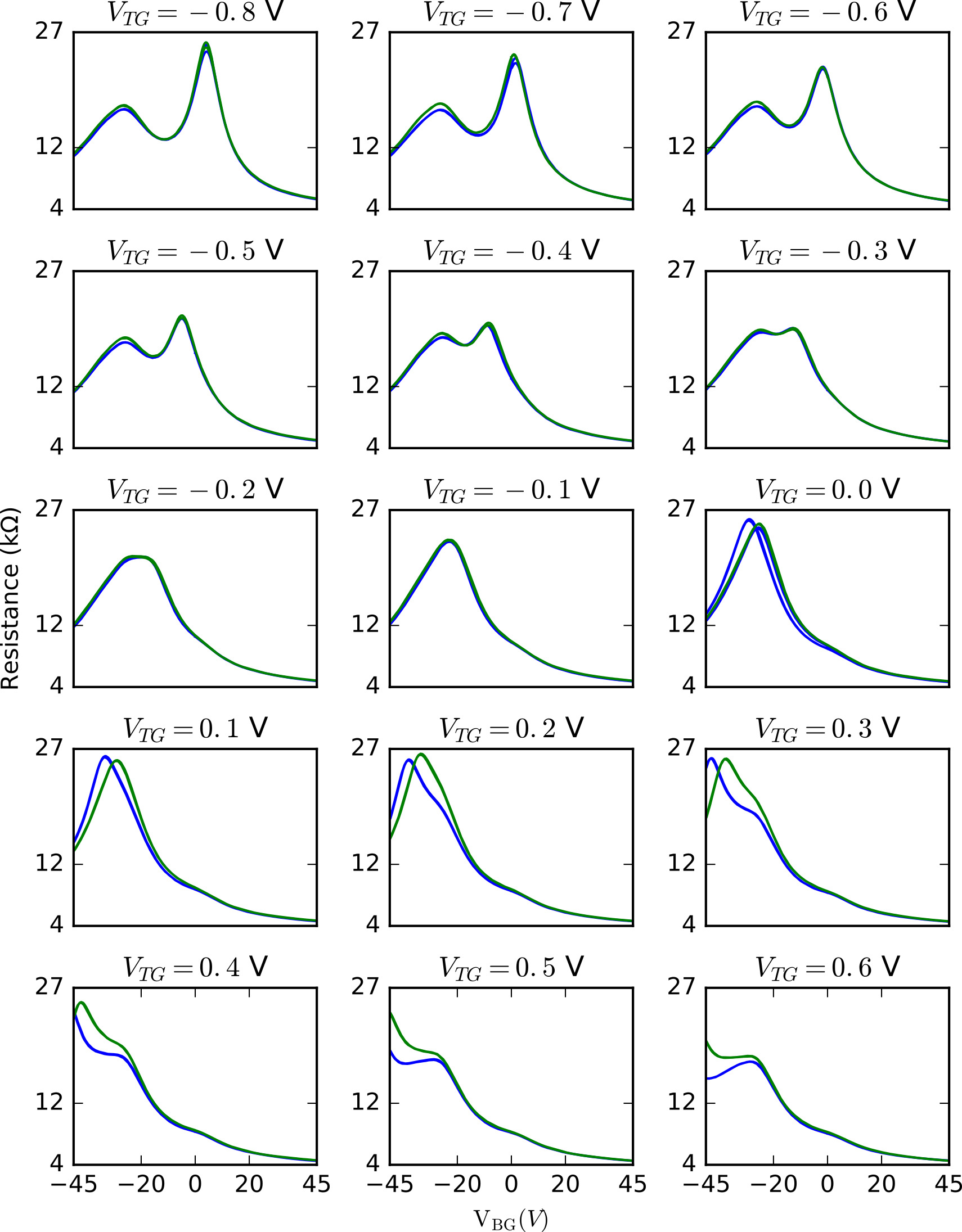}
\caption{Resistance at 60 K measured before and after 15 thermal
  cycling events.}
\label{fig:thermal2}
\end{figure}

\begin{figure}
    \centering
\includegraphics[width=150mm]{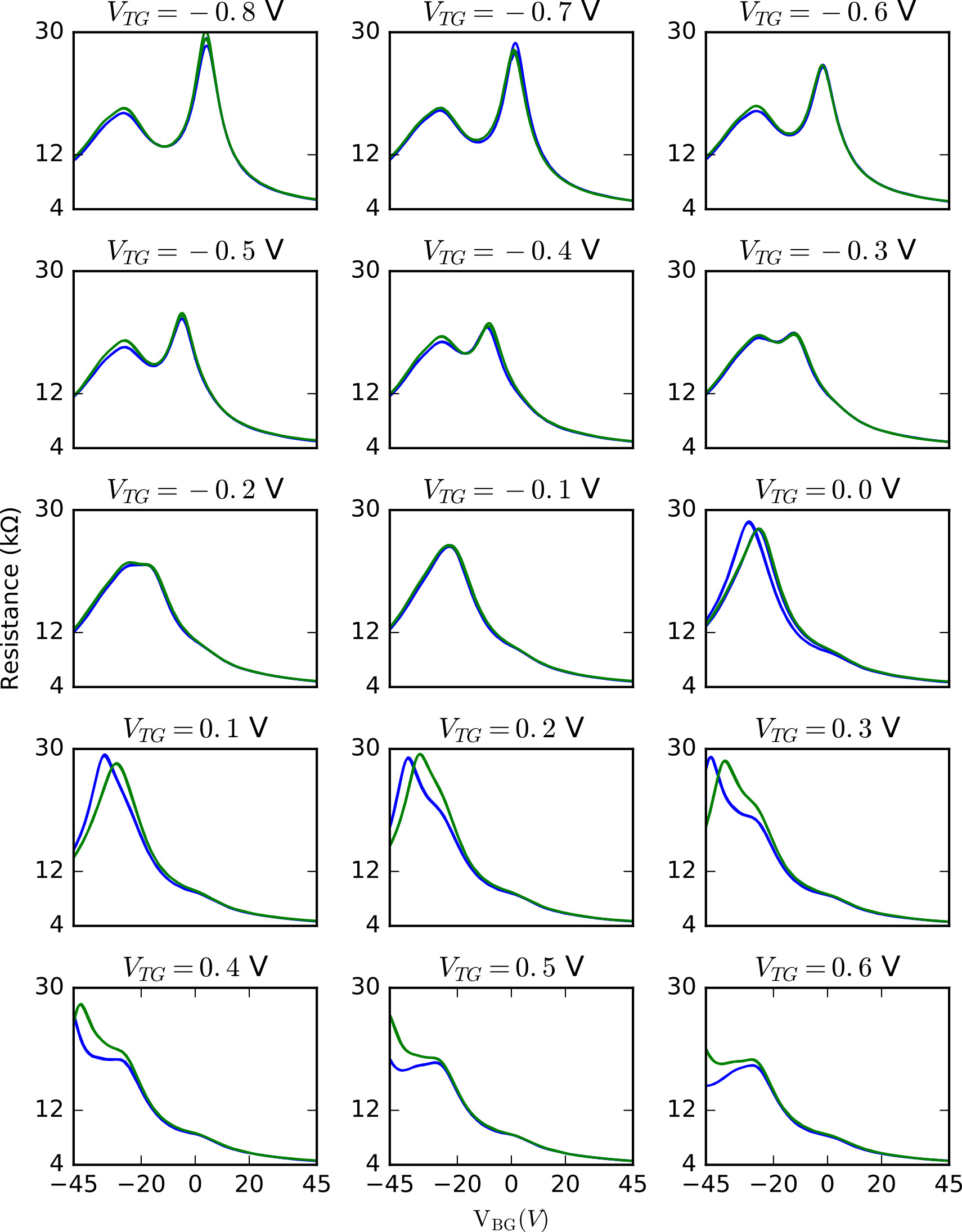}
\caption{Resistance at 30 K measured before and after 15 thermal
  cycling events.}
\label{fig:thermal3}
\end{figure}

We have verified that the \textit{p-n} junction formed using the
technique demonstrated in the paper is not affected by thermal
cycling. We have measured both the resistance and photovoltage with
before and after 15 thermal cycles between measurements taken at the
same top gate voltage.  Figures~\ref{fig:thermal1}, \ref{fig:thermal2}
and \ref{fig:thermal3} show plots of resistance taken at three
different temperatures – 30, 60 and 120 K for different top gate
voltages. The two plots are plotted in blue and green and there are 15
thermal cycling between them. All plots show both forward and backward
sweeps of the gate voltage so that hysteresis, if present, can be
identified.  At all three temperatures, the gating curves overlap
indicating that the p-n junction created is immune to thermal cycling
effects.

\begin{figure}
  \centering
  \includegraphics[width=150mm]{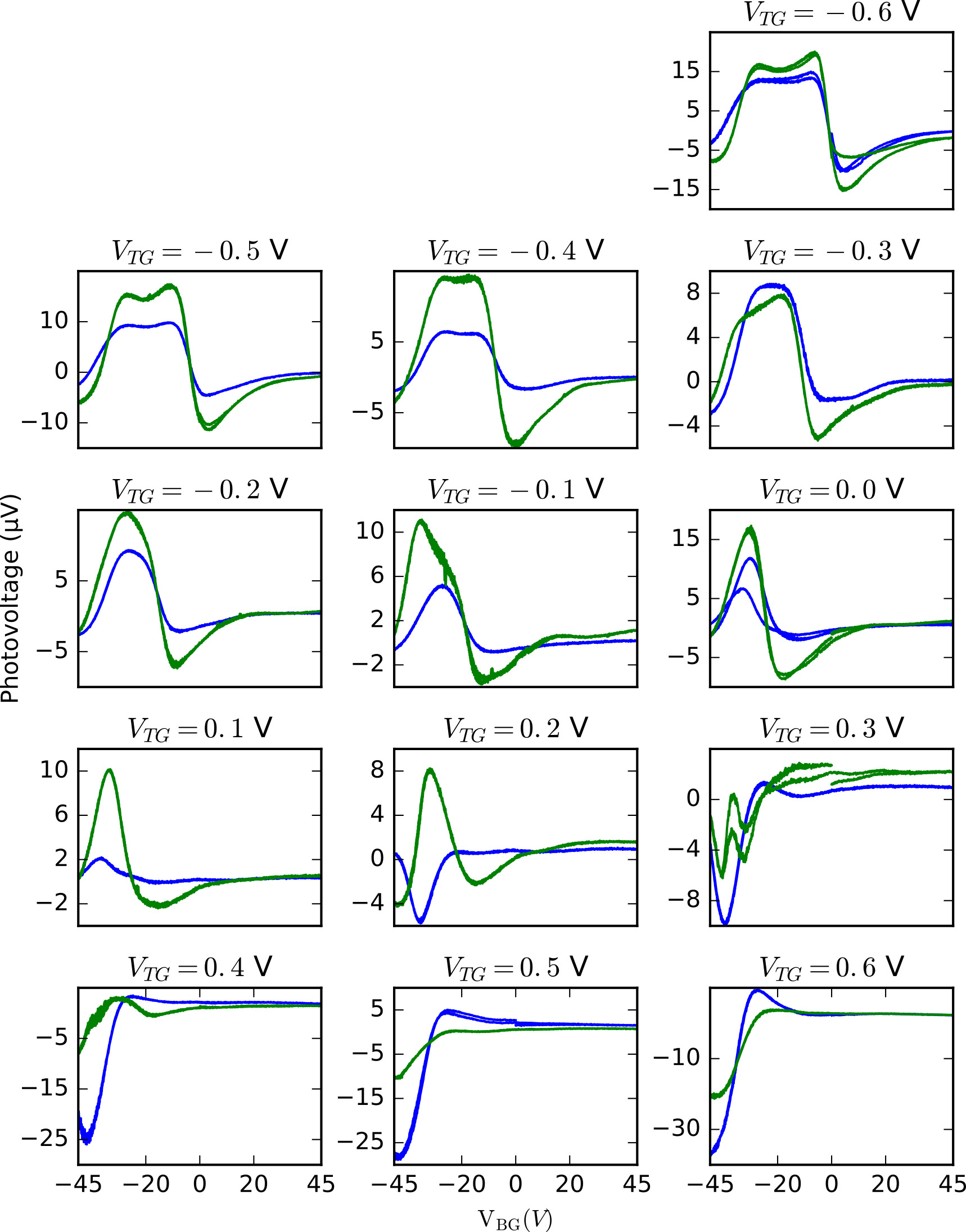}
\caption{Photovoltage at 60 K measured before and after 15 thermal
  cycling events.}
\label{fig:thermal4}
\end{figure}

\begin{figure}
  \centering
  \includegraphics[width=150mm]{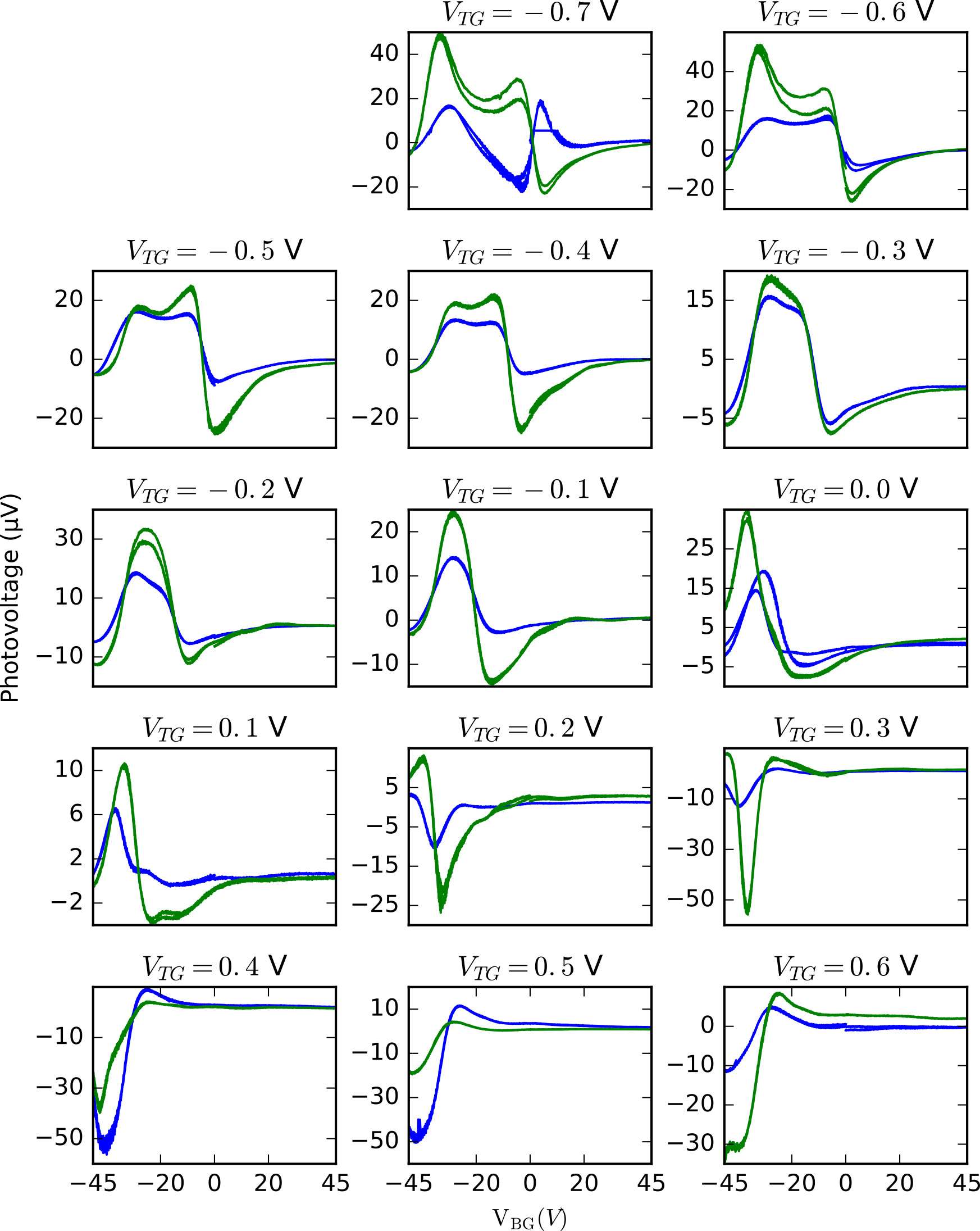}
\caption{Photovoltage at 30 K measured before and after 15 thermal
  cycling events.}
\label{fig:thermal5}
\end{figure}

Similarly, the photovoltage has also been measured at 30 K and 60 K
and is shown in Figures~\ref{fig:thermal4}~and~\ref{fig:thermal5} ,
with 15 thermal cycling events between the curves. Here, the
difference between the two measurement st is more pronounced. However,
the trend is clearly the same. Each thermal cycling event takes a
significant amount of time – nearly 4 hours – including the time we
wait for the sample to stabilize. During this time, any position
change of the sample due to thermal drift in the sample stage, or any
long- term drift in the z-position of microscope objective that is
used to focus light – even by a few microns - will result in a change
in the intensity and position of light illumination. This can account
for the changes we have observe in the photovoltage cycling events
shown below.

We noticed a decrease in the top gate capacitance during the first
heating and cooling cycle where it decreased from $90\, C_{BG}$ to
$40 \, C_{BG}$ ($C_{BG}$: back gate capacitance) and subsequently
remained stable over the course of measurements which involved more
than 50 thermal cycles.

\clearpage

\section{Temperature Dependence}

\begin{figure}[h!]
\centering
\includegraphics[width=70mm]{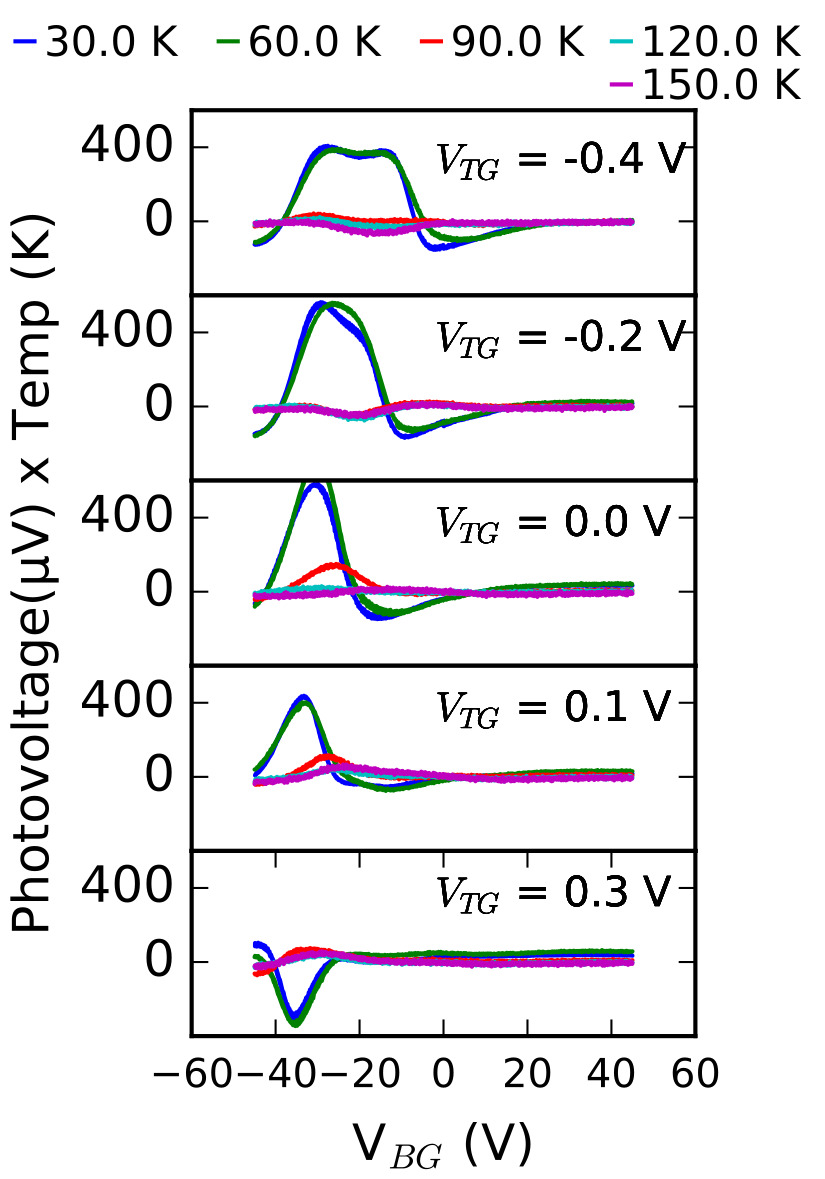}
\caption{Temperature dependence of the photovoltage shown by plotting
  the product of the photovoltage and temperature at different top
  gate voltages.}
\label{fig:thermal1B}
\end{figure}

In Figure~\ref{fig:thermal1B}, we have plotted the product of the
photovoltage and temperature as a function of the back gate voltage
for different top gate voltages. Since the 30 K and 60 K curves
overlap, the photovoltage $PV \propto 1/T $ in this temperature range.

\clearpage


\end{document}